\documentclass[11pt]{article}

\usepackage{amsmath}
\usepackage{amsfonts}
\makeatletter \@addtoreset{equation}{section} \makeatother

\newcommand{\wh}{\widehat}

\newcommand{\cD}{{\cal D}}
\newcommand{\fD}{\mathfrak{D}}

\newcommand{\cF}{{\cal F}}
\newcommand{\cG}{{\cal G}}
\newcommand{\cH}{{\cal H}}
\newcommand{\cL}{{\cal L}}
\newcommand{\cM}{{\cal M}}
\newcommand{\cN}{{\cal N}}

\newcommand{\cP}{{\cal P}}
\newcommand{\cQ}{{\cal Q}}
\newcommand{\cR}{{\cal R}}

\newcommand{\cV}{{\cal V}}

\newcommand{\tI}{{\tilde I}}
\newcommand{\tJ}{{\tilde J}}

\newcommand{\la}{\lambda}
\newcommand{\si}{\sigma}

\def\a{\alpha}
\def\b{\beta}
\def\g{\gamma}
\newcommand{\bea}{\begin{eqnarray}}
\newcommand{\eea}{\end{eqnarray}}
\newcommand{\nnu}{\nonumber}
\newcommand{\be}{\begin{equation}}
\newcommand{\ee}{\end{equation}}
\newcommand{\bt}{\begin{tabular}}
\newcommand{\et}{\end{tabular}}
\newcommand{\ba}{\begin{array}}
\newcommand{\ea}{\end{array}}
\newcommand{\bbm}{\begin{bmatrix}}
\newcommand{\ebm}{\end{bmatrix}}
\newcommand{\lra}{\longrightarrow}
\newcommand{\Lra}{\Longrightarrow}
\newcommand{\prt}{\partial}
\newcommand{\Dsl}{D\negthickspace\negthickspace\negthinspace /}
\newcommand{\prtsl}{\partial\negthickspace\negthickspace /}
\newcommand{\calDsl}{\mathcal{D}\negthickspace\negthickspace\negthinspace /}
 
\newcommand{\smf}[2]{{\textstyle\frac{#1}{#2}}}
\newcommand{\gs}{g_{\scriptscriptstyle S}}
\newcommand{\ga}{g_{\scriptscriptstyle A}}
\newcommand{\uA}{{\scriptscriptstyle (A)}}
\newcommand{\uS}{{\scriptscriptstyle (S)}}
\def\remark#1{{}}
\def\eqn#1{(\ref{#1})}

\let\LARGE=\Large
\let\Large=\large
\begin{document}

%%%%%%%%%%%%%%%%%%%%%%%%%%%%%%%%%%%%%%%%%%%%%%%%%%%%%%

\thispagestyle{empty}
\rightline{HU-EP 01/09}
\rightline{hep-th/0103106}

\vspace{8truemm}

\centerline{\bf \LARGE
 General Matter Coupled 
$\cN=4$ Gauged Supergravity}
\bigskip

\centerline{\bf \LARGE in Five Dimensions}

\bigskip

\vspace{1.2truecm}

\centerline{\bf G. Dall'Agata\footnote{e-mail:
dallagat@physik.hu-berlin.de},  
 C. Herrmann\footnote{e-mail: 
 herrmann@physik.uni-halle.de} and \ 
M. Zagermann\footnote{e-mail: 
 zagermann@physik.uni-halle.de}}

\vspace{.5truecm}
\centerline{\em $^1$Institut f\"ur Physik, Humboldt-Universit\"at
Berlin}
\centerline{\em Invalidenstrasse 110, D-10115 Berlin, Germany}
\medskip

\centerline{ \em $^{2,3}$Fachbereich Physik,
Martin-Luther-Universit\"at
Halle-Wittenberg}
\centerline{\em Friedemann-Bach-Platz 6, D-06099 Halle, Germany}

\vspace{1truecm}

%%%%%%%%%%%%%%%%%%%%%%%%%%%%%%%%%%%%%%%%%%%%%%%%%%%%%%%%

\begin{abstract}
  We construct the general form of matter coupled $\cN=4$ gauged supergravity
  in five dimensions.  Depending on the structure of the gauge group, these
  theories are found to involve vector and/or tensor multiplets. When
  self-dual tensor fields are present, they must be charged under a
  one-dimensional Abelian group and cannot transform non-trivially under any
  other part of the gauge group. A short analysis of the possible ground
  states of the different types of theories is given. It is found that AdS
  ground states are only possible when the gauge group is a direct product of
  a one-dimensional Abelian group and a semi-simple group. In the purely
  Abelian, as well as in the purely semi-simple gauging, at most Minkowski
  ground states are possible.  The existence of such Minkowski ground
  states could be proven in the compact Abelian case.
\end{abstract}

\vfill

March 2001~.

%%%%%%%%%%%%%%%%%%%%%%%%%%%%%%%%%%%%%%%%%%%%%%%%%%%%%%%%%%%%%%%%

\newpage

%%%%%%%%%%%%%%%%%%%%%%%%%%%%%%%%%%%%%%%%%%%%%%%%%%%%%%%%%%%%%%%%%

\section{Introduction}

%%%%%%%%%%%%%%%%%%%%%%%%%%%%%%%%%%%%%%%%%%%%%%%%%%%%%%%

The last few years have witnessed a renewed intense interest in
five-dimensional gauged supergravity theories. This interest was largely
driven by the study of the AdS/CFT corres\-pondence 
\cite{402,GKP98,406,401}, but
also by recent attempts to construct a supersymmetric version of the
Randall-Sundrum (RS) scenario \cite{050,090,110,888,271,272,273,2732,274,275}.

One especially fruitful direction in the study of the AdS/CFT correspondence
was its generalization to include certain four-dimensional quantum field
theories with non-trivial renormalization group
(RG) flows.  The best-studied
examples of such RG-flows arise from relevant perturbations of the
$d=4,\mathcal{N}=4$ super Yang-Mills (SYM) theory, and were mapped to domain
wall solutions of $d=5,\mathcal{N}=8$ gauged supergravity (see
\cite{351,352,FGPW99} for the first explicit examples).

In \cite{FGPW99}, an RG-flow interpolating between two supersymmetric
conformal fixed points of a mass deformed version of $d=4,\cN=4$ SYM was
studied. The corresponding domain wall solution was constructed after a
truncation of the $d=5,\mathcal{N}=8$ supergravity theory to a particular
$\mathcal{N}=4$ subsector. This subsector describes $\mathcal{N}=4$
supergravity coupled to two $\mathcal{N}=4$ tensor multiplets with scalar
manifold $\mathcal{M}=SO(1,1)\times SO(5,2)/[SO(5)\times SO(2)]$ and gauge
group $SU(2)\times U(1)$.  In the same paper this particular $\mathcal{N}=4$
theory was also conjectured to be the holographic dual of the common sector of
all $d=4,\mathcal{N}=2$ superconformal gauge theories based on ``quiver''
diagrams \cite{DM96,LNV98}. The holographic duals of these quiver gauge
theories were identified in refs.\ \cite{403,LNV98} as IIB string theory on
$AdS_5 \times (S^5 /\Gamma)$, where $\Gamma$ is a discrete subgroup of
$SU(2)\subset SU(4)$ of the ADE type.

The Kaluza-Klein spectrum of IIB string theory on $AdS_5 \times (S^5 /\Gamma)$
was studied in ref.\ \cite{411}, where the IIB supergravity modes were fit
into $SU(2,2|2)$ multiplets, and in ref.\ \cite{G98}, which also considered
the $\Gamma=\mathbb{Z}_n$-twisted string states.  These twisted states live on
$AdS_5\times S^1$, and were found to correspond to five-dimensional,
$\mathcal{N}=4$ ``self-dual'' tensor multiplets of $SU(2,2|2)$ \cite{G98}.

It has therefore been suggested \cite{GNS99,FGPW99} that five-dimensional,
$\mathcal{N}=4$ gauged supergravity coupled to vector and/or tensor multiplets
might encode some non-trivial information on the $\mathcal{N}=2$ quiver
theories even though a \emph{ten}-dimensional supergravity description of the
corresponding orbifold theory is not available. In particular, it has been
conjectured in \cite{GNS99} that certain aspects of the flow from the
$S^5/{\mathbb{Z}}_2$ to the $T^{1,1}$ compactification of IIB string theory
\cite{KW98} might be captured by a $1/4$ BPS domain wall solution of a
suitable $\mathcal{N}=4$ gauged supergravity theory with tensor multiplets.
The tensor multiplets are crucial for such a flow because they host the
supergravity states dual to the twisted gauge theory operators that drive this
flow \cite{GNS99}.

Unfortunately, too little was known about five-dimensional, $\mathcal{N}=4$
gauged supergravity coupled to an arbitrary number of matter (i.e., vector-
and tensor-) multiplets in order to further investigate the corresponding
gravity description.  In fact, only a $SU(2)\times U(1)$ gauging of pure
$\mathcal{N}=4$ supergravity \cite{RO86} and a peculiar $SU(2)$ gauging of
$\mathcal{N}=4$ supergravity coupled to vector multiplets \cite{AT85} were
available so far.

It is the purpose of this paper to close this gap in the literature and to
construct the general matter-coupled five-dimensional, $\mathcal{N}=4$ gauged
supergravity.

This, with \cite{080,GZ99,GZ00b,CD00} for the $\cN = 2$ case and
\cite{GRW86,PPN85} for the $\cN = 8$ case, completes also the description of
all standard gauged supergravity theories in $d = 5$.

These theories should also help to clarify whether the different no-go
theorems that have been put forward against a \emph{smooth}
supersymmetrization of the RS scenario \cite{090,110,888,999,275} can be
extended to the $\mathcal{N}=4$ sector as well. As for the (more successful)
supersymmetrizations based on singular brane sources
\cite{271,272,273,2732,274}, the theories constructed in this paper allow for
the possibility to confine $\mathcal{N}=2$ supergravity theories on the brane,
as it is the self-dual tensor fields in the bulk that were shown to circumvent
the ``no photons on the brane'' theorems \cite{DLS00,CLP00}.

The paper is organized as follows. All the gauged $\mathcal{N}=4$ supergravity
theories we will construct can be derived from the ungauged Maxwell/Einstein
supergravity theories (MESGT) studied in \cite{AT85}.  Section \ref{gen_set},
therefore, first recalls the relevant properties of these ungauged theories.
In Section \ref{glo_sym}, then, we will take a closer look at the global
symmetries of these theories and analyze to what extent these global
symmetries can be turned into local gauge symmetries. This discussion will
reveal some interesting differences to the $\mathcal{N}=2$ and the
$\mathcal{N}=8$ cases and also results in a rather natural way to organize the
rest of the paper: Section \ref{KA} discusses the general gauging of an
Abelian group, which turns out to require the introduction of tensor
multiplets. In Section \ref{KAKS}, we will then construct the combined gauging
of a semi-simple group and an Abelian group. The resulting scalar potentials
are then briefly analyzed in Section \ref{sec_pot}, which also contains the
reductions to various relevant special cases previously considered in the
literature.  In the last section, we draw some conclusions and list a few
 open problems.

%%%%%%%%%%%%%%%%%%%%%%%%%%%%%%%%%%%%%%%%%%%%%%%%%%%%%%%%%
\section{Ungauged Maxwell/Einstein supergravity}
%%%%%%%%%%%%%%%%%%%%%%%%%%%%%%%%%%%%%%%%%%%%%%%%%%%%%%%%%

\subsection{General setup}
\label{gen_set}

The starting point of our construction is the ungauged MESGT of ref.\ 
\cite{AT85} which describes the coupling of $n$ vector multiplets to $\cN=4$
supergravity\footnote{For ungauged five-dimensional supergravity theories,
  vector and tensor fields are Poincar\'e dual, and we therefore do not have
  to distinguish between vector and tensor multiplets at this level.}. Our
spacetime conventions coincide with those of ref.\ \cite{AT85} and are further
explained in Appendix A.  

The field content of the $\cN=4$ supergravity
multiplet is
\begin{equation}
\Big(\,\,e_\mu{}^m\,,~\psi_\mu^i\,,~A_\mu^{ij}\,,~a_\mu\,,
~\chi^i\,,~\sigma\,\Big),
\label{n=4_sugra}
\end{equation}
i.e., it contains one graviton $e_\mu{}^m$, four gravitini $\psi_\mu^i$, six
vector fields $(A_\mu^{ij},a_\mu)$, four spin $1/2$ fermions $\chi^i$ and one
 real scalar field $\sigma$. 
Here, $\mu/m$ are the five-dimensional Einstein/Lorentz indices and
 the indices $i,j=1,\ldots,4$ correspond to the fundamental
representation of the $R$-symmetry group $USp(4)$ of the underlying $\cN=4$
Poincar\'e superalgebra. The vector field $a_\mu$ is $USp(4)$ inert, whereas
the vector fields $A_\mu^{ij}$ transform in the {\bf 5} of $USp(4)$, i.e.,
\begin{equation}
A_\mu^{ij} \ = \ -A_\mu^{ji}~,\qquad A_\mu^{ij}\,\Omega_{ij} \ = \ 0,
\end{equation}
with $\Omega_{ij}$ being the symplectic metric of $USp(4)$. In the following,
we will sometimes
make use of the local isomorphism
$SO(5)\cong USp(4)$ to denote $SO(5)$ representations using $USp(4)$
indices.

An $\cN=4$ vector
multiplet is given by
\begin{equation}
\Big(\,A_\mu\,,~\lambda^i\,,~\phi^{ij}\,\Big),
\end{equation}
where $A_\mu$ is a vector field, $\lambda^i$ denotes four spin $1/2$ fields,
and the $\phi^{ij}$ are scalar fields transforming in the {\bf 5} of
$USp(4)$.  

Coupling $n$ vector multiplets to supergravity, the
field content of the theory can be summarized as follows
\begin{equation}
\Big(\,e_\mu{}^m\,,~\psi_\mu^i\,,~A_\mu^\tI\,,~a_\mu\,,~\chi^i\,,~\la^{ia}\,,~
\si\,,~\phi^x\,\Big).
\label{fullcontent}
\end{equation}
Here, $a=1,\ldots,n$ counts
the number of vector multiplets whereas
 $\tI=1,\ldots,(5+n)$  collectively denotes
the $A_\mu^{ij}$ and the vector fields of the vector multiplets. Similarly, 
$x=1,\ldots,5n$ is a collective index for the scalar fields in the vector
multiplets. 
We will further adopt the following convention to raise and lower $USp(4)$
indices: 
\begin{equation}
T^i \ = \ \Omega^{ij}\,T_j~,\quad T_i \ = \ T^j\,\Omega_{ji},
\end{equation}
whereas $a,b$ are raised and lowered with $\delta^{ab}$.

As was shown in \cite{AT85}, the
manifold spanned by the $(5n+1)$ scalar fields is
\begin{equation}
    \label{cM}
\cM \ = \ \frac{SO(5,n)}{SO(5)\times SO(n)}\times SO(1,1),
\end{equation}
where the $SO(1,1)$ part corresponds to the $USp(4)$-singlet $\sigma$
of the supergravity multiplet.
The theory has therefore a {\em global} symmetry group 
$SO(5,n)\times SO(1,1)$ and a {\em local composite}
$SO(5)\times SO(n)$ invariance.
  
The coset part of the scalar manifold $\cM$
can be described in two ways: one can either, as in (\ref{fullcontent}), 
choose a
parameterization in terms of coordinates $\phi^x$, where
the metric $g_{xy}$ on the coset part of $\cM$ 
is given in terms of $SO(5)\times SO(n)$
vielbeins by (cf.\ Appendix B):
\begin{equation}
g_{xy} \ = \ \frac14\,f_x^{ija}\,f^a_{yij},
\end{equation}
such that the kinetic term for the scalar fields takes the typical form
$\frac12 g_{xy}\prt_\mu \phi^{x}\prt^\mu \phi^y$
 of a
 non-linear sigma-model. This way of describing $\cM$ is particularly useful
 for discussing geometrical properties of the theory.

As an alternative description, one can use coset representatives 
$\,L_\tI{}^A\,$ 
where $\tI$ denotes a $\cG=SO(5,n)\,$ 
index, and $A=(ij,a)$ is a $\cH=SO(5)\times SO(n)\,$ index. 
Denoting the inverse of $\,L_\tI{}^A\,$ by $\,L_A{}^\tI\,$ (i.e., 
$\,L_\tI{}^A\, \,L_B{}^\tI\ = \delta_{B}^{A}$), the 
vielbeins on $\cG/\cH$ and the composite $\cH$-connections 
are determined from the 1-form:
\begin{equation}
L^{-1}dL \ = \ Q^{ab}\,\mathfrak{T}_{ab} + Q^{ij}\,
\mathfrak{T}_{ij} + P^{aij} \, \mathfrak{T}_{aij},
\end{equation}
where $\,(\mathfrak{T}_{ab},\mathfrak{T}_{ij})\,$ are the
generators of the Lie algebra $\mathfrak{h}$
of $\cH$, and $\mathfrak{T}_{aij}$ denotes the  generators of the coset part of
the 
Lie algebra $\mathfrak{g}$ of $\cG$. More precisely, 
\begin{equation}\label{Q}
Q^{ab} \ = \ L^{\tI a}dL_\tI{}^b\qquad\mathrm{and}\qquad Q^{ij} \ = \
L^{\tI ik} 
dL_{\tI k}{}^j
\end{equation}
are the composite $SO(n)$ and $USp(4)$ connections, respectively,  and
\begin{equation}\label{P}
P^{aij} \ = \  L^{\tI a} 
dL_{\tI}{}^{ij}=-\frac{1}{2}f_{x}^{aij}\,d\phi^x
\end{equation}
describes the space-time pull-back of the $\cG/\cH$ vielbein.
Note that $Q_\mu^{ab}$ is antisymmetric in the
$SO(n)$ indices, whereas $Q_\mu^{ij}$ is symmetric in $i$ and $j$.
This second way of parameterizing the scalar manifold is particularly useful
to exhibit the action of the different invariance groups as clearly as
possible. We will make this choice in what follows 
for the construction of the gauged theories. Eqs.\ (\ref{P}) and
(\ref{diff1})-(\ref{diff4}) can be used to easily
switch between both formalisms. Appendix B contains more details on the
geometry of $\cG/\cH$.

The Lagrangian of the
ungauged MESGT  reads \cite{AT85}:
\bea
e^{-1}\,\cL &=& -\frac12 R -\frac12\,
\bar\psi_\mu^i\,\Gamma^{\mu\nu\rho}\,D_\nu\psi_{\rho
  i}-\frac14\Sigma^2\,a_{\tI\tJ}\,F_{\mu\nu}^\tI
F^{\mu\nu\tJ}-\frac14\Sigma^{-4}\,G_{\mu\nu}G^{\mu\nu}\nnu\\
&&-\frac12\,(\prt_\mu\si)^2-\frac12\,\bar\chi^i\,
\Dsl\chi_i-\frac12\bar\la^{ia}\Dsl\la_i^a-\frac12\,P_\mu^{aij}P^\mu_{aij}\nnu\\
&&-\frac{i}{2}\bar\chi^i\Gamma^\mu\Gamma^\nu\psi_{\mu
  i}\,\prt_\nu\si
+i\bar\la^{ia}\,\Gamma^\mu\Gamma^\nu\psi_\mu^j\,P_{\nu ij}{}^a\nnu\\
&&{}+\frac{\sqrt 3}{6}\Sigma
L_\tI^{ij}\,F_{\rho\si}^\tI\bar\chi_i\,\Gamma^\mu\Gamma^{\rho\si}\psi_{\mu
  j}-\frac14\Sigma L_\tI^a\,F_{\rho\si}^\tI\bar\la^{ai}\,\Gamma^\mu
\Gamma^{\rho\si}\psi_{\mu i}\nnu\\ 
&&-\frac{1}{2\sqrt6}\Sigma^{-2}\,\bar\chi^i\,\Gamma^\mu\Gamma^{\rho\si}\psi_{\mu
  i}G_{\rho\si}+\frac{5i}{24\sqrt
  2}\Sigma^{-2}\bar\chi^i\Gamma^{\rho\si}\chi_i G_{\rho\si}\nnu\\
&&-\frac{i}{12}\Sigma L_\tI^{ij}
F_{\rho\si}^\tI\bar\chi_i\,\Gamma^{\rho\si}\chi_j -\frac{i}{2\sqrt
3}\Sigma
L_\tI^a\,F_{\rho\si}^\tI\bar\la^{ia}\Gamma^{\rho\si}\chi_i-\frac{i}{8\sqrt
  2}\Sigma^{-2} G_{\rho\si}\bar\la^{ia}\Gamma^{\rho\si}\la_i^a\nnu\\
&&+\frac{i}{4}\Sigma L_\tI^{ij}
F_{\rho\si}^\tI\bar\la^a_i\Gamma^{\rho\si}\la_j^a-\frac{i}{4}\Sigma
L_\tI^{ij} F_{\rho\si}^\tI\,[\bar\psi_{\mu
i}\Gamma^{\mu\nu\rho\si}\psi_{\nu
  j}+2\,\bar\psi^\rho_i\,\psi^\si_j]\nnu\\
&&-\frac{i}{8\sqrt2}\,\Sigma^{-2} G_{\rho\si}\,[\bar\psi^i_{\mu}
\Gamma^{\mu\nu\rho\si}\psi_{\nu
  i}+2\bar\psi^{\rho i}\,\psi^\si_i]\nnu\\ 
&&{}+\frac{\sqrt2}{8}e^{-1}\,C_{\tI\tJ}\,\epsilon^{\mu\nu\rho\si\la}\,
F_{\mu\nu}^\tI F_{\rho\si}^\tJ\,a_\la +e^{-1}\cL_{4\!f},
\label{Lagrange1}
\eea
and the supersymmetry transformation laws are given by\footnote{We use the 
following (anti)
symmetrization
symbols:$(ij)\equiv\smf12 (ij+ji)~,~[ij]\equiv\smf12 (ij-ji)$.}
\bea
\delta e_{\mu}^{m}&=&\frac{1}{2}\bar\varepsilon^{i}
\Gamma^{m}\psi_{\mu i}\nonumber\\
\delta \psi_{\mu i}&=& D_{\mu}\varepsilon_{i} + 
\frac{i}{6}\Sigma L_{\tI ij} F_{\rho \sigma}^{\tI} 
(\Gamma_{\mu}^{\,\,\rho \sigma}
-4 \delta^{\rho}_{\mu}\Gamma^{\sigma}) \varepsilon^{j}\nonumber\\
&&+\frac{i}{12\sqrt{2}} \Sigma^{-2}G_{\rho \sigma}
(\Gamma_{\mu}^{\,\,\rho \sigma}
-4 \delta^{\rho}_{\mu}\Gamma^{\sigma})\varepsilon_{i} +
\textrm{3-fermion terms}\nonumber\nnu\\
\delta \chi_{i}&=&-\frac{i}{2}\prtsl \sigma
\varepsilon_{i}+\frac{\sqrt{3}}{6}
\Sigma L_{\tI ij} F_{\rho \sigma}^{\tI} \Gamma^{\rho
\sigma}\varepsilon^{j}
-\frac{1}{2\sqrt{6}}\Sigma^{-2}G_{\rho \sigma} \Gamma^{\rho \sigma}
\varepsilon_{i}\nonumber\\
\delta \lambda^{a}_{i}&=&iP_{\mu ij}^{a}
\Gamma^{\mu} \varepsilon^{j} -\frac{1}{4}\,\Sigma\,
L_{\tI}^{a} F_{\rho \sigma}^{\tI} \Gamma^{\rho \sigma}\varepsilon_{i}
  +
\textrm{3-fermion terms}\nonumber\\
\delta A_{\mu}^{\tI}&=&\vartheta_{\mu}^{\tI}\nonumber\\
\delta a_{\mu}&=&\frac{1}{\sqrt{6}}\Sigma^{2}{\bar{\varepsilon}}^{i}
\Gamma_{\mu}\chi_{i}-\frac{i}{2\sqrt{2}}\Sigma^{2}{\bar{\varepsilon}}^{i}
\psi_{\mu i}\nonumber\\
\delta \sigma&=&\frac{i}{2}{\bar{\varepsilon}}^{i}\chi_{i}\nonumber\\
\delta
L_{\tI}^{ij}&=&-iL_{\tI}^{a}(\delta^{[i}_{k}\delta^{j]}_{l}-\frac{1}{4}
\Omega^{ij}\Omega_{kl}){\bar{\varepsilon}}^{k}\lambda^{la}\nonumber\\
\delta L_{\tI}^{a}&=&-iL_{\tI ij}{\bar{\varepsilon}}^{i}\lambda^{ja}
\label{trafo1}
\eea
with 
\begin{equation}\label{theta}
\vartheta^{\tI}_{\mu}\equiv -\frac{1}{\sqrt{3}}\Sigma^{-1}L^{\tI}_{ij}
{\bar{\varepsilon}}^{i}\Gamma_{\mu}\chi^{j}-i\Sigma^{-1}L^{\tI}_{ij}
{\bar{\varepsilon}}^{i}\psi^{j}_{\mu}+\frac{1}{2}L^{\tI}_{a}\Sigma^{-1}
{\bar{\varepsilon}}^{i}\Gamma_{\mu}\lambda^{a}_{i}.
\end{equation}

Here,
\begin{equation}
F_{\mu\nu}^\tI\ = (\prt_\mu A_\nu^\tI-\prt_\nu A_\mu^\tI)~,\quad
G_{\mu\nu}\ = (\prt_\mu a_\nu-\prt_\nu a_\mu)~,
\end{equation}
are the Abelian field strengths of the vector fields, whereas
the scalar field in the supergravity
multiplet is parameterized by
\begin{equation}
\Sigma \ = \ e^{\frac{1}{\sqrt 3}\si}.
\end{equation}
Moreover, the covariant derivative, $D_{\mu}$, that acts on the
fermions is given by
\begin{equation}
D_\mu\la_{i}^a \ = \ \nabla_\mu\la_{i}^a+Q_{\mu i}{}^j\la_{j}^a+Q_{\mu}^{ab}
\la_{i}^b,
\end{equation}
with $\nabla_\mu$ being the Lorentz- and spacetime covariant
derivative. 

Supersymmetry imposes
\begin{equation}
a_{\tI\tJ} \ = \ L_\tI^{ij}L_{\tJ ij}+L_\tI^aL_\tJ^a~,\quad C_{\tI\tJ}
\ = \
L_\tI^{ij}L_{\tJ ij}-L_\tI^aL_\tJ^a,
\end{equation}
where $a_{\tI\tJ}$ acts as a metric on the $\tI,\tJ$ indices:
\begin{equation}
L_{\tI}^A\ = \ a_{\tI\tJ}\,L^{\tJ A}.
\end{equation}
The symmetric tensor $C_{\tI\tJ}$ turns out to be
 {\em constant}, and in fact is nothing but the $SO(5,n)$ metric.

 Supersymmetry also imposes a number of additional algebraic and differential
 relations on the $L_{\tI}^{A}$, which are listed in Appendix B.  Actually,
 the requirement of invariance under supersymmetry heavily constrains the form
 of the possible couplings and, as we show in Appendix C using the superspace
 language, if one does not consider non-minimal couplings, those presented
 here already use all the freedom allowed by the $\cN = 4$ superalgebra.

\subsection{The global symmetries and their possible gaugings}
\label{glo_sym}

The above ungauged Maxwell/Einstein supergravity theories are subject to
several global and local invariances.  Apart from supersymmetry, the
\emph{local} invariances are\footnote{In large parts of the literature on
  ``gauged'' supergravity, none of these two local invariances is referred to
  as a ``gauge'' symmetry: the local composite $Usp(4)\times SO(n)$ symmetry
  is not based on physical vector fields, and the Maxwell-type invariance is,
  in this context, more viewed as a special case of a more general invariance
  under $C^{(p)}\longrightarrow C^{(p)}+d\Lambda^{(p-1)}$ for arbitrary
  $p$-form fields $C^{(p)}$ and $(p-1)$-forms $\Lambda^{(p-1)}$.  The term
  ``gauged'' supergravity, by contrast, is used when some physical
  vector fields of a (usually $\mathcal{N}$-extended) supergravity theory
  couple to other matter fields via gauge covariant derivatives.  In most
  cases, the gauge symmetry of such a theory reduces to a global symmetry of
  an ``ungauged'' supergravity theory when the gauge coupling is turned off.}
   
\begin{itemize}
\item a local composite $USp(4)\times SO(n)$ symmetry 
\item a Maxwell-type invariance of the form\footnote{Note that it is essential
    for this symmetry to hold that the $C_{\tI\tJ}$ be constant.}
\begin{eqnarray}\label{Maxwell}
A^{\tI}_{\mu} &\longrightarrow& A^{\tI}_{\mu} +\prt_{\mu}
\Lambda^{\tI}\\
a_{\mu} &\longrightarrow& a_{\mu} +\prt_{\mu} \Lambda.
\end{eqnarray}

\end{itemize}

In addition, the ungauged
theories of Section \ref{gen_set} are invariant under \emph{global}
 $SO(1,1)\times SO(5,n)$ transformations,  
which exclusively
act on the vector fields, the coset representatives $L^{\tI}_{A}$,  and
the 
scalar field $\Sigma=e^{\frac{\sigma}{\sqrt{3}}}$ 
according to
\bea
\delta_{SO(5,n)} A_{\mu}^{\tI}&=&
\alpha^{r}M_{(r)\tJ}^{\tI} A_{\mu}^{\tJ}\nonumber\\
\delta_{SO(5,n)} L_{A}^{\tI}&=&
\alpha^{r}M_{(r)\tJ}^{\tI} L_{A}^{\tJ}\nonumber\\
\delta_{SO(5,n)} \Sigma &=& 0\nonumber\\
\delta_{SO(5,n)} a_{\mu}&=&0\label{SO(5,n)trafo}
\eea
and
\bea
\delta_{SO(1,1)} A_{\mu}^{\tI}&=&-\lambda A_{\mu}^{\tI}\nonumber\\
\delta_{SO(1,1)} L_{A}^{\tI}&=&0\nonumber\\
\delta_{SO(1,1)} \Sigma&=&\lambda \Sigma\nonumber\\
\delta_{SO(1,1)} a_{\mu}&=&2\lambda a_{\mu},
\eea
where $M_{(r)\tJ}^{\tI}\in \mathfrak{so}(5,n)$, and $\alpha^{r}$ 
$(r=1,\ldots,\dim(SO(5,n)))$
and $\lambda$ are some infinitesimal parameters. 

We will now analyze to what extent this \emph{global} $SO(1,1)\times
SO(5,n)$
invariance can be turned into a \emph{local} 
(i.e., Yang-Mills-type) gauge
symmetry by introducing minimal couplings of some of the vector fields,
a process commonly referred to as ``gauging''.

The first thing to notice is that the $SO(1,1)$ factor can\emph{not} be
gauged, as \emph{all} the vector fields transform non-trivially under it
(i.e., none of the vector fields in the theory could serve as the
corresponding (neutral) Abelian gauge field). Thus, we can restrict ourselves
to gaugings of subgroups $K\subset SO(5,n)$.

The vector fields $A_{\mu}^{\tI}$ and the coset representatives
$L^{\tI}_{A}$ transform in the $\mathbf{(5+n)}$
of $SO(5,n)$; all other fields are $SO(5,n)$-inert
 (cf.\ eqs.\ (\ref{SO(5,n)trafo})). Hence, any gauge group $K\subset
SO(5,n)$
has to act non-trivially on at least some of the vector fields 
$A_{\mu}^{\tI}$ (as well as  on some of the coset representatives $\,L_A^\tI$).

Having physical
applications in mind, we will restrict our subsequent discussion
  to gauge groups
$K$ that are either
\begin{enumerate}
\item Abelian or
\item semi-simple or
\item a direct product of a semi-simple and an Abelian  group.
\end{enumerate}

\subsubsection{$K$ is Abelian}
\label{sec_KA}

Let us first assume that $K$ is Abelian. As we mentioned above, $K$ has
to
act non-trivially on at least some of the vector fields
$A_{\mu}^{\tI}$.
In addition to such non-singlets of $K$,
there might also be vector fields among the $A_{\mu}^{\tI}$ that are
$K$-inert.
Thus, in general, we have a decomposition of the form
\begin{equation}
\mathbf{(5+n)}_{SO(5,n)}\longrightarrow \textrm{singlets}(K) \oplus 
\textrm{non-singlets}(K).
\end{equation}
Let us split the vector fields $A_{\mu}^{\tI}$ accordingly:
\begin{displaymath}
A_{\mu}^{\tI}=(A_{\mu}^{I},A_{\mu}^{M})
\end{displaymath}
where the indices $I,J,\ldots$ label the $K$-singlets, and
$M,N,\ldots$ denote the non-singlets. 

As was first pointed out in \cite{GRW86,PPN85} for the $\mathcal{N}=8$
theory, the presence of such non-singlet vector fields poses a 
problem for the supersymmetric gauging of $K$ and requires the dualization of
the charged vectors into self-dual \cite{TPN84} tensor fields.

As we will explain in Section 3, 
this conversion of the $A_{\mu}^{M}$
to ``self-dual'' tensor fields $B_{\mu\nu}^{M}$ is achieved in practice
by simply replacing all field strengths  $F_{\mu\nu}^{M}$ by
$B_{\mu\nu}^{M}$
and by adding a kinetic term of the form $\mathcal{L}_{BdB}=B\wedge dB$
to the Lagrangian. The derivative in $\mathcal{L}_{BdB}$ then
turns out to be  automatically
$K$-covariantized by the $B\wedge B\wedge a$-term in the Lagrangian,
which 
originates from the $F\wedge F \wedge a$ term
in the ungauged Lagrangian (\ref{Lagrange1})  upon the replacement 
$F_{\mu\nu}^{M}\longrightarrow B_{\mu\nu}^{M}$. This has an important
consequence: only the $SO(5,n)$ singlet vector field $a_{\mu}$
can couple to the tensor fields $B_{\mu\nu}^{M}$ for this kind of
gauging,
and, consequently, any Abelian gauge group $K$ can be at most
one-dimensional,
i.e., it can be either $SO(2)$ or $SO(1,1)$ with gauge vector $a_\mu$. 
Note that the converse is
also 
true: If tensor fields have to be introduced in order to gauge a 
(not necessarily Abelian) subgroup
$K\subset SO(5,n)$, these tensor fields can only be charged with
respect to a one-dimensional Abelian subgroup of $K$.
This is an interesting difference to the $\mathcal{N}=8$ and the
$\mathcal{N}=2$ theories, where the tensor fields
can also transform in nontrivial representations of 
non-Abelian gauge groups $K$ \cite{GRW86,PPN85,GZ99}.

We will discuss the gauging of an Abelian group $K$ in Section \ref{KA}.

\subsubsection{$K$ is semi-simple}
\label{sec_KS}

Let us now come to the case when the gauge group $K\subset SO(5,n)$ is
semi-simple. In that case, some of the vector fields $A_{\mu}^{\tI}$ of the
ungauged theory have to transform in the adjoint representation of $K$ so that
they can be promoted to the corresponding Yang-Mills-type gauge fields. Put
another way, the $\mathbf{(5+n)}$ of $SO(5,n)$ has to contain the adjoint of
$K$ as a sub-representation. A priori, one would
therefore expect the decomposition
\begin{equation}
\mathbf{(5+n)}_{SO(5,n)}\longrightarrow \textrm{adjoint}(K)\oplus
\textrm{singlets}(K)\oplus \textrm{non-singlets}(K).
\end{equation}

Just as for the Abelian case, any non-singlet vector fields outside the
adjoint of $K$ would have to be converted to ``self-dual'' tensor fields
$B_{\mu\nu}^{M}$. On the other hand, we already found that, due to the
peculiar structure of the Chern-Simons term in the ungauged theory
(\ref{Lagrange1}), only the vector field $a_{\mu}$ could possibly couple to
such tensor fields. As one vector field alone can never gauge a semi-simple
group, we are led to the conclusion that the gauging of a semi-simple group
$K$ can \emph{not} introduce any tensor fields, and we can restrict ourselves
to the case
\begin{equation}\label{decomposition3}
\mathbf{(5+n)}_{SO(5,n)}\longrightarrow \textrm{adjoint}(K)\oplus
\textrm{singlets}(K).
\end{equation}
The vector fields in the adjoint will then serve as the $K$ gauge
fields,
whereas the singlets are mere ``spectator'' vector fields.

Let us conclude this subsection with a rough classification of the
possible \emph{compact} semi-simple gauge groups $K$.
Obviously, a  compact gauge group $K$ has to be a subgroup of the
maximal 
compact subgroup, $SO(5)\times SO(n)\subset SO(5,n)$. Furthermore,
$K$ 
can be either a subgroup of the $SO(5)$ factor or a subgroup of the
$SO(n)$
factor or a direct product of both of these. 

If $K\subset SO(5)$,
it can only be the standard $SO(3)\cong SU(2)$ subgroup of $SO(5)$, 
because this is the only semi-simple subgroup of $SO(5)$ with the
property
that the $\mathbf{5}$ of $SO(5)$ decomposes according to 
(\ref{decomposition3}).

If $K\subset SO(n)$, the fundamental representation of $SO(n)$ has to contain
the adjoint of $K$ (as well as possibly some $K$-singlets).
However, the adjoint of \emph{any} compact semi-simple group $K$ 
can be embedded into the fundamental representation of $SO(n)$
as long as $\dim(K)\leq n$ (One simply has to take 
the positive definite Cartan-Killing
metric as the $SO(\dim(K))$ metric). Thus, \emph{any}
compact  semi-simple group $K$ can be gauged along the above lines
as long as $\dim(K)\leq n$.

An obvious combination of the previous two paragraphs finally
covers the case when $K$ is a direct product of an 
$SO(5)$- and an $SO(n)$-subgroup.

We will not write down the gauging of a semi-simple group separately, as it
can be easily obtained as a special case of the combined Abelian and
semi-simple gauging, which we discuss now.

\subsubsection{$K=K_{\textrm{Abelian}}\times K_{\textrm{semi-simple}}$}
\label{sec_KAKS}

If $K$ is a direct product of a semi-simple and an Abelian group, 
one simply has to combine Sections \ref{sec_KA} and \ref{sec_KS}:

Let $K_{S}$ denote the semi-simple and $K_{A}$ the Abelian part of $K$. Then
the gauging of $K_{A}$ implies
\begin{equation}
\mathbf{(5+n)}_{SO(5,n)}\longrightarrow \textrm{singlets}(K_{A})\oplus
\textrm{non-singlets}(K_{A}).
\label{200}
\end{equation}
As discussed in Section \ref{sec_KA}, the non-singlet vector fields of $K_A$
have to be converted to self-dual tensor fields. These tensor fields cannot be
charged under the semi-simple part $K_S$ (see the discussion of Section
\ref{sec_KS}). Hence, $K_S$ can only act on the singlets of
$K_A$. Furthermore, the action of $K_S$ on these $K_A$
singlets cannot introduce additional tensor fields, i.e., there can be no
non-singlets of $K_S$ beyond the adjoint of $K_S$:
\begin{equation}
\textrm{singlets}(K_{A})\longrightarrow
\textrm{adjoint}(K_{S})\oplus
\textrm{singlets}(K_{S}).
\label{201}
\end{equation}

As described in Section \ref{sec_KA}, the 
Chern-Simons term of the ungauged theory (\ref{Lagrange1}) requires 
$K_{A}$ to be one-dimensional,
i.e., we can have either $K=U(1)\times K_{S}$ or $K=SO(1,1)\times K_{S}$.

We will take a closer look at this general gauging in Section \ref{KAKS}.

%%%%%%%%%%%%%%%%%%%%%%%%%%%%%%%%%%%%%%%%%%%%%%%%%%%%%%%%%%%%%%%%%%%
\section{The gauging of an Abelian group $K_{A}\subset SO(5,n)$}
\label{KA}
%%%%%%%%%%%%%%%%%%%%%%%%%%%%%%%%%%%%%%%%%%%%%%%%%%%%%%%%%%%%%%%%%%%%%

The goal of this paper is to construct the most general gauging of $\cN=4$
supergravity coupled to an arbitrary number of vector and tensor multiplets.
As was discussed in the previous section, this corresponds to the simultaneous
gauging of a one-dimensional Abelian group $K_A$ and a semi-simple group
$K_S$.  However, it is the Abelian gauging that introduces the tensor fields,
so it is worth treating this technically more subtle case separately:

The gauging of an Abelian group $K_{A}\subset SO(5,n)$ 
proceeds in three steps (see also the $\cN=2$ \cite{GZ99} and $\cN=8$
\cite{GRW86,PPN85} cases):

\noindent\textbf{Step 1:}\\
As discussed in Section \ref{sec_KA}, the most general decomposition
of the $\mathbf{(5+n)}$ of $SO(5,n)$ with respect to an Abelian gauge
group
 $K_{A}$ is
\begin{equation}
\mathbf{(5+n)}_{SO(5,n)}\longrightarrow 
\textrm{singlets}(K_{A})\oplus \textrm{non-singlets}(K_{A}),
\end{equation}
We will again use $I,J,\ldots$ for the $K_{A}$ singlets and
$M,N,\ldots$
for the non-singlets of $K_{A}$. Note that the (rigid) $K_{A}$
invariance
of the Chern Simons term in (\ref{Lagrange1}) already
implies that we have $C_{IM}=0$: If
$C_{IM}\neq 0$, we would need $\Lambda_{N}^{M}C_{IM}=0$ for the invariance
of $C_{IM}F^{I}\wedge F^{M}\wedge a$, where
$\Lambda_{N}^{M}$ denotes the $K_{A}$ transformation matrix of the
non-singlet
vector fields $A_{\mu}^{M}$. But then the $K_{A}$ representation space
spanned
 by the $A_{\mu}^{M}$ would have to have one non-trivial
null-eigenvector,
i.e., there would be at least one singlet among the $A_{\mu}^{M}$,
contrary to our assumption. 
This same condition $C_{IM}=0$ also follows by requiring the closure 
of the supersymmetry algebra on the vector fields $A_{\mu}^{I}$, which in the
superspace analysis amounts to the requirement of the closure of the 
Bianchi Identities for the supercurvatures $F^I$.
 
In order to gauge $K_{A}$, the non-singlet vector fields
$A_{\mu}^{M}$ now have to be converted to ``self-dual'' tensor fields 
$B_{\mu\nu}^{M}$, whereas the singlet vector 
fields $A_{\mu}^{I}$ will play the r\^{o}le
of spectator vector fields.
Using the N\oe ther procedure, 
the tensor field ``dualization'' is done by first
literally replacing all $F_{\mu\nu}^{M}$ in the ungauged Lagrangian
(\ref{Lagrange1}) and the ungauged transformation laws (\ref{trafo1})
by tensor fields $B_{\mu\nu}^{M}$. This is more than a mere 
relabeling, as the $B_{\mu\nu}^{M}$ are no longer assumed to be
the curls of vector fields $A_{\mu}^{M}$. Because of this, we no longer
have a Bianchi identity for the  tensor fields $B_{\mu\nu}^{M}$,
i.e., in general, we now have $dB^{M}\neq 0$. 
This, however, already breaks supersymmetry, because some of the
supersymmetry variations in the ungauged theory only vanished due to
 $dF^{M}=0$. As a remedy, one therefore adds the extra term
\begin{equation}
\mathcal{L}_{BdB}=
\frac{1}{4 \ga}\epsilon^{\mu\nu\rho\sigma\lambda}\Omega_{MN}
B_{\mu\nu}^{M}\partial_{\rho}B_{\sigma\lambda}^{N}
\end{equation}
to the Lagrangian and
requires the supersymmetry transformation rule of the $B_{\mu\nu}^{M}$
to be
\bea\label{deltaB}
\delta B_{\mu\nu}^{M}&=&2 \partial_{[\mu}\vartheta_{\nu]}^{M}
+2\ga a_{[\mu}\Lambda_{N}^{M}\vartheta_{\nu]}^{N}
-\ga \Sigma
L_{Nij}\Omega^{NM}\bar\psi_{[\mu}^{i}\Gamma_{\nu]}\varepsilon^{j}\nnu\\
&&+\frac{i}{4}\ga \Sigma
L_{N}^{a}\Omega^{NM}\bar\lambda^{ia}\Gamma_{\mu\nu}
\varepsilon_{i}-\frac{i}{2\sqrt{3}}\ga \Sigma
L_{Nij}\Omega^{NM}\bar\chi^{i}
\Gamma_{\mu\nu}\varepsilon^{j}.
\eea
Here, $\ga $ denotes the $K_{A}$ coupling constant, $\Omega_{MN}$
is a constant and antisymmetric metric with inverse $\Omega^{MN}$:
\begin{displaymath}
\Omega_{MN}\Omega^{NP}=\delta_{M}^{P}, \qquad \Omega_{MN}=-\Omega_{NM},
\end{displaymath}
and $\vartheta_{\mu}^{M}$ is as defined in eq.\ (\ref{theta}).
It is not too hard  to show  that inserting $\delta B_{\mu\nu}^{M}$
into
$\mathcal{L}_{BdB}$ exactly cancels the supersymmetry breaking terms
that arise
due to $dB^{M}\neq 0$.

\noindent\textbf{Step 2:}\\
Closer inspection of $\mathcal{L}_{BdB}$ reveals that it 
combines with 
\begin{displaymath}
\mathcal{L}_{BBa}=\frac{\sqrt{2}}{8}C_{MN}
\epsilon^{\mu\nu\rho\sigma\lambda}B_{\mu\nu}^{M}B_{\rho\sigma}^{N}
a_{\lambda}
\end{displaymath}
(which stems from the  former Chern-Simons term in (\ref{Lagrange1})),
to automatically form a $K_{A}$-covariant derivative:
\begin{displaymath}
\mathcal{L}_{BdB}+\mathcal{L}_{BBa}=\frac{1}{4 \ga }
\epsilon^{\mu\nu\rho\sigma\lambda}\Omega_{MN}
B_{\mu\nu}^{M}\mathfrak{D}_{\rho}B_{\sigma\lambda}^{N},
\end{displaymath}
where 
\begin{equation}\label{covB}
\mathfrak{D}_{\rho}B_{\mu\nu}^{M}\equiv \nabla_{\rho}B_{\mu\nu}^{M}+
\ga a_{\rho}\Lambda^{M}_{\,N}B_{\mu\nu}^{N}
\end{equation}
with
\begin{equation}\label{Lambda}
\Lambda^{M}_{\,N}=\frac{1}{\sqrt{2}}\Omega^{MP}C_{PN}
\end{equation}
being the $K_{A}$-transformation matrix of the tensor fields.
The same is true for the first two terms in $\delta B_{\mu\nu}^{M}$
(eq.\ (\ref{deltaB})),
which naturally combine to $2\mathfrak{D}_{[\mu}\vartheta_{\nu]}^{M}$.
However, apart from these two derivatives, none of the various other
derivatives in the Lagrangian or the transformation laws are 
covariantized with respect to $K_{A}$. On the other hand, the only
derivatives
affected by this are derivatives acting on  the coset representatives
$L_{M}^{A}$, $L_{A}^{M}$, which exclusively appear inside the composite
connections $Q_{\mu i}^{\,\,\,\,\,\,j}$, $Q_{\mu a}^{\,\,\,\,\,\,b}$
and the $P_{\mu ij}^{a}$ (cf.\ eqs.\ (\ref{Q}), (\ref{P})). 
Introducing $K_{A}$-covariant derivatives
\begin{equation}
\mathfrak{D}_{\rho}L_{A}^{M}\equiv \partial_{\rho}L_{A}^{M}+
\ga a_{\rho}\Lambda^{M}_{N}L_{A}^{N}
\end{equation}
also inside $Q_{\mu i}^{\,\,\,\,\,\,j}$, $Q_{\mu a}^{\,\,\,\,\,\,b}$
and  $P_{\mu ij}^{a}$ is tantamount to making the
replacements
\bea
Q_{\mu i}^{\,\,\,\, \,\,j}&\longrightarrow&\mathcal{Q}_{\mu
i}^{\,\,\,\,\,\,j}
=Q_{\mu i}^{\,\,\,\,\,\,j}-\ga 
a_{\mu}\Lambda^{M}_{N}L_{Mik}L^{Nkj}\label{Q2}\\
Q_{\mu a}^{\,\,\,\,\,\, b}&\longrightarrow&
\mathcal{Q}_{\mu a}^{\,\,\,\,\,\, b}=Q_{\mu a}^{\,\,\,\,\,\, b}+\ga 
a_{\mu}\Lambda^{M}_{N}L_{Ma}L^{Nb}\label{Q3}\\
P_{\mu ij}^{a}&\longrightarrow&\mathcal{P}_{\mu ij}^{a}=
P_{\mu ij}^{a}-\ga 
a_{\mu}\Lambda^{M}_{N}L_{Mij}L^{Na}
\eea
everywhere in the Lagrangian and the transformation laws.

\noindent\textbf{Step 3:}\\
After Step 1 and Step 2, the theory is now $K_{A}$ gauge invariant
to all orders in $\ga $
 and supersymmetric to order $(\ga )^{0}$.
However, at order $(\ga )^{\geq 1}$ the theory fails to be
supersymmetric 
due to the numerous new $\ga$-dependent terms we introduced.
As a third step, one therefore has to restore supersymmetry by
adding a few  
$\ga $-dependent (but gauge invariant) terms
to the covariantized Lagrangian and transformation rules.

After all these modifications, the
final Lagrangian is given by (up to 4-Fermion terms):

\bea
e^{-1}\,\cL &=& -\frac{1}{2} R -\frac{1}{2}\,
\bar\psi_\mu^i\,\Gamma^{\mu\nu\rho}\,\mathcal{D}_\nu\psi_{\rho
  i}-\frac{1}{4}\Sigma^2\,a_{\tI\tJ}\,H_{\mu\nu}^\tI
H^{\mu\nu\tJ}-\frac{1}{4}\Sigma^{-4}\,G_{\mu\nu}G^{\mu\nu}\nnu\\
&&-\frac{1}{2}\,(\prt_\mu\si)^2-\frac{1}{2}\,\bar\chi^i\,
\calDsl\chi_i-\frac{1}{2}
\bar\la^{ia}\calDsl\la_i^a-\frac{1}{2}\,\mathcal{P}_\mu^{aij}
\mathcal{P}^\mu_{aij}\nnu\\
&&-\frac{i}{2}\bar\chi^i\Gamma^\mu\Gamma^\nu\psi_{\mu
  i}\,\prt_\nu\si
+i\bar\la^{ia}\,\Gamma^\mu\Gamma^\nu\psi_\mu^j\,\mathcal{P}_{\nu
ij}{}^a\nnu\\
&&{}+\frac{\sqrt 3}{6}\Sigma
L_\tI^{ij}\,H_{\rho\si}^\tI\bar\chi_i\,\Gamma^\mu\Gamma^{\rho\si}\psi_{\mu
  j}-\frac{1}{4}\Sigma L_\tI^a\,H_{\rho\si}^\tI\bar\la^{ai}\,\Gamma^\mu
\Gamma^{\rho\si}\psi_{\mu i}\nnu\\ 
&&-\frac{1}{2\sqrt6}\Sigma^{-2}\,\bar\chi^i\,
\Gamma^\mu\Gamma^{\rho\si}\psi_{\mu
  i}G_{\rho\si}+\frac{5i}{24\sqrt
  2}\Sigma^{-2}\bar\chi^i\Gamma^{\rho\si}\chi_i G_{\rho\si}\nnu\\
&&-\frac{i}{12}\Sigma L_\tI^{ij}
H_{\rho\si}^\tI\bar\chi_i\,\Gamma^{\rho\si}\chi_j -\frac{i}{2\sqrt
3}\Sigma
L_\tI^a\,H_{\rho\si}^\tI\bar\la^{ia}\Gamma^{\rho\si}\chi_i-\frac{i}{8\sqrt
  2}\Sigma^{-2} G_{\rho\si}\bar\la^{ia}\Gamma^{\rho\si}\la_i^a\nnu\\
&&+\frac{i}{4}\Sigma L_\tI^{ij}
H_{\rho\si}^\tI\bar\la^a_i\Gamma^{\rho\si}\la_j^a-\frac{i}{4}\Sigma
L_\tI^{ij} H_{\rho\si}^\tI\,[\bar\psi_{\mu
i}\Gamma^{\mu\nu\rho\si}\psi_{\nu
  j}+2\,\bar\psi^\rho_i\,\psi^\si_j]\nnu\\
&&-\frac{i}{8\sqrt2}\,\Sigma^{-2} G_{\rho\si}\,[\bar\psi^i_{\mu}
\Gamma^{\mu\nu\rho\si}\psi_{\nu
  i}+2\bar\psi^{\rho i}\,\psi^\si_i]\nnu\\ 
&&{}+\frac{\sqrt2}{8}e^{-1}\,C_{IJ}\,\epsilon^{\mu\nu\rho\si\la}\,
F_{\mu\nu}^I F_{\rho\si}^J\,a_\la 
+\frac{e^{-1}}{4 \ga }\epsilon^{\mu\nu\rho\sigma\lambda}\Omega_{MN}
B_{\mu\nu}^{M}\mathfrak{D}_{\rho}B_{\sigma\lambda}^{N}\nnu\\
&&{}+\frac{3i}{2}\ga U_{ij}\bar\psi^{i}_{\mu}\Gamma^{\mu\nu}
\psi_{\nu}^{j} + i \ga N_{ijab}\bar\lambda^{ia}\lambda^{jb} -
\frac{5i}{2}\ga U_{ij}\bar\chi^{i}\chi^{j}\nnu\\
&&-\ga V_{ij}^{a}\bar\psi^{i}_{\mu}\Gamma^{\mu}\lambda^{ja}-2\sqrt3 
\ga U_{ij}
\bar\psi^{i}_{\mu}\Gamma^{\mu}\chi^{j}-\frac{4i}{\sqrt{3}}\ga V_{ij}^{a}
\bar\chi^{i}\lambda^{ja}-\ga ^{2}\cV^\uA, 
\label{Lagrange2}
\eea

and the supersymmetry transformation laws are given by (up to 3-fermion
terms)
\bea
\delta e_{\mu}^{m}&=&\frac{1}{2}\bar\varepsilon^{i}
\Gamma^{m}\psi_{\mu i}\nonumber\\
\delta \psi_{\mu i}&=& \mathcal{D}_{\mu}\varepsilon_{i} + 
\frac{i}{6}\Sigma L_{\tI ij} H_{\rho \sigma}^{\tI} 
(\Gamma_{\mu}^{\,\,\rho \sigma}
-4 \delta^{\rho}_{\mu}\Gamma^{\sigma}) \varepsilon^{j}\nonumber\\
&&+\frac{i}{12\sqrt{2}} \Sigma^{-2}G_{\rho \sigma}
(\Gamma_{\mu}^{\,\,\rho \sigma}
-4 \delta^{\rho}_{\mu}\Gamma^{\sigma})\varepsilon_{i}
+i \ga U_{ij}\Gamma_{\mu}\varepsilon^{j}\nonumber\\
\delta \chi_{i}&=&-\frac{i}{2}\prtsl \sigma
\varepsilon_{i}+\frac{\sqrt{3}}{6}
\Sigma L_{\tI ij} H_{\rho \sigma}^{\tI} \Gamma^{\rho
\sigma}\varepsilon^{j}
-\frac{1}{2\sqrt{6}}\Sigma^{-2}G_{\rho \sigma} \Gamma^{\rho \sigma}
\varepsilon_{i}-2\sqrt3 \ga U_{ij}\varepsilon^{j}\nonumber\\
\delta \lambda^{a}_{i}&=&i\mathcal{P}_{\mu ij}^{a}\Gamma^{\mu} 
\varepsilon^{j} -\frac{1}{4}\,\Sigma
L_{\tI}^{a} H_{\rho \sigma}^{\tI} \Gamma^{\rho \sigma}\varepsilon_{i}
 +\ga  V_{ij}^{a}\varepsilon^{j}\nonumber\\
\delta A_{\mu}^{I}&=&\vartheta_{\mu}^{I}\nonumber\\
\delta B_{\mu\nu}^{M}&=&2
\mathfrak{D}_{[\mu}\vartheta_{\nu]}^{M}-\ga \Sigma
L_{Nij}\Omega^{NM}\bar\psi_{[\mu}^{i}\Gamma_{\nu]}\varepsilon^{j}\nnu\\
&&+\frac{i}{4}\ga \Sigma
L_{N}^{a}\Omega^{NM}\bar\lambda^{ia}\Gamma_{\mu\nu}
\varepsilon_{i}-\frac{i}{2\sqrt{3}}\ga \Sigma
L_{Nij}\Omega^{NM}\bar\chi^{i}
\Gamma_{\mu\nu}\varepsilon^{j}\nnu\\
\delta a_{\mu}&=&\frac{1}{\sqrt{6}}\Sigma^{2}{\bar{\varepsilon}}^{i}
\Gamma_{\mu}\chi_{i}-\frac{i}{2\sqrt{2}}\Sigma^{2}{\bar{\varepsilon}}^{i}
\psi_{\mu i}\nonumber\\
\delta \sigma&=&\frac{i}{2}{\bar{\varepsilon}}^{i}\chi_{i}\nonumber
\eea
\bea
\delta
L_{\tI}^{ij}&=&-iL_{\tI}^{a}(\delta^{[i}_{k}\delta^{j]}_{l}-\frac{1}{4}
\Omega^{ij}\Omega_{kl}){\bar{\varepsilon}}^{k}\lambda^{la}\nonumber\\
\delta L_{\tI}^{a}&=&-iL_{\tI ij}{\bar{\varepsilon}}^{i}\lambda^{ja}.
\label{trafo2}
\eea
In the above expressions, we have introduced the tensorial quantity
\begin{equation}
H_{\mu\nu}^{\tI}\equiv (F_{\mu\nu}^{I},B_{\mu\nu}^{M})
\end{equation}
as well as the new $USp(4)\times SO(n)$ covariant
derivatives
\begin{equation}
\mathcal{D}_{\mu}\lambda^{a}_{i}=\nabla_{\mu}\lambda^{a}_{i}
+\mathcal{Q}_{\mu
i}^{\,\,\,\,\,\,j}\lambda_{j}^{a}+\mathcal{Q}_{\mu}^{ab}
\lambda^{b}_{i}
\label{der_ferm}
\end{equation}
with the new, $\ga $-dependent connections (\ref{Q2}) and (\ref{Q3}).

The new scalar field dependent quantities $U_{ij}$, $V_{ij}^{a}$,
$N_{ijab}$,
as well as the scalar potential $\cV^\uA$ are fixed by supersymmetry:
\bea
U_{ij}&=& \frac{\sqrt2}{6}\Sigma^{2}\Lambda^{N}_{M}L_{N(i|k|}
L^{Mk}{}_{j)}\\
V_{ij}^{a}&=&\frac{1}{\sqrt{2}}\Sigma^{2}\Lambda^{N}_{M}L_{Nij}
L^{Ma}\\
N_{ijab}&=&-\frac{1}{2\sqrt{2}}\Sigma^{2}\Lambda^{N}_{M}L_{N}^{a}
L^{Mb}\Omega_{ij}+\frac{3}{2}U_{ij}\delta_{ab}\\
\cV^\uA&=&\frac{1}{2}V_{ij}^{a}V^{aij}.
\eea

In particular, the $U_{ij}$ and $V_{ij}^a$ shifts in the 
supersymmetry transformations of the fermionic fields are all 
expressed in terms of the so-called ``boosted structure constants'', which 
are just the representation matrices of the vector fields under the 
gauged group multiplied by the coset representatives.
This is a common feature of all gauged supergravities which have 
a scalar manifold given by an ordinary homogeneous space   
\cite{Castellani:1986ka}.

%%%%%%%%%%%%%%%%%%%%%%%%%%%%%%%%%%%%%%%%%%%%%%%%%%%%%%%%%%
\section{The simultaneous gauging of $K_{S}\times K_{A}$}
\label{KAKS}
%%%%%%%%%%%%%%%%%%%%%%%%%%%%%%%%%%%%%%%%%%%%%%%%%%%%%%%%%%

The starting point for the simultaneous gauging of $K_S\times K_A$ 
is the $K_A$ gauged
Lagrangian (\ref{Lagrange2}) with the corresponding supersymmetry variations
(\ref{trafo2}).
In this case, 
fields carrying an $SO(5,n)$ index $\tI$ decompose according to
(\ref{200}) and (\ref{201}) into adjoint($K_S$), singlets($K_S$) and
non-singlets($K_A$) fields. 
However, to avoid the introduction of a third type of
$SO(5,n)$ index beside $I$ and $M$, we will collectively denote 
the adjoint($K_S$) and singlet($K_S$) fields by
$I,J,K$, while non-singlet($K_A$) fields will carry indices $M,N$. 
The actual distinction between the
adjoint($K_S$) and the singlet($K_S$) fields 
will be made implicitly by the $K_S$
structure constant $f_{IJ}^K$ which will be taken  to 
vanish whenever one index denotes a $K_S$ singlet field. In order for the
Chern-Simons term to be globally 
$K_S$ invariant, these structure constants have to
satisfy the following identity
\begin{equation}
C_{IJ}\,f^I_{KL} + C_{IL}\,f^I_{KJ} \ = \ 0.
\end{equation}

The additional gauging of $K_S$ essentially proceeds in two steps:

\noindent
\textbf{Step 1:}\\
All derivatives acting on $K_S$ charged fields must be $K_S$ covariantized.
This modifies the definition of the field strengths of the gauge fields
$A_\mu^I$ in the
standard way: 
\bea F^I_{\mu\nu}& \lra& \cF^I_{\mu\nu} \ = \ F^I_{\mu\nu}
+\gs\,A_\mu^J
f_{JK}^I A_\nu^K\nnu\\ 
\mathrm{thus}\quad
H_{\mu\nu}^\tI &\lra& \cH^\tI_{\mu\nu} \ = \ (\cF^I_{\mu\nu}, B_{\mu\nu}^M).
\eea 
In order to simplify the notation, we use the same derivative symbols as
in the previous section, but now also with
 $\gs$ dependent contributions:
\begin{equation}
\fD_\mu L^\tI_A \ = \ \prt_\mu L^\tI_A +\ga\delta_M^\tI
a_\mu\Lambda^M_N 
L^N_A +\gs\delta^\tI_I A_\mu^J f_{JK}^I L^K_A.
\end{equation}
This in turns modifies the $USp(4)$ and $SO(n)$ connections, as well as
the vielbein $\,\cP_{\mu ij}^a\,$, as these quantities
 contain derivatives of the coset representatives. 
Again, we use the same symbols as in the previous
section, but now also include the new $\gs$ dependent contributions, i.e.,
\bea
\cQ_{\mu i}^{\ \ \ j} &=& Q_{\mu i}^{\ \ \ j} -\ga a_\mu\Lambda^M_N
L_{Mik}L^{Nkj}
+\gs A_\mu^J L^K_{ik} f_{JK}^I L_{I}^{kj},\label{compco1}\\
\cQ_{\mu a}^{\ \ \ b} &=& Q_{\mu a}^{\ \ \ b} +\ga
a_\mu\Lambda^M_NL_{Ma}L^{Nb}
-\gs A_\mu^J L^K_a f_{JK}^I L_{I}^b,\label{compco2}\\
\cP_{\mu ij}^a &=& P_{\mu ij}^a -\ga a_\mu\Lambda^M_N L_{Mij}L^{Na}-\gs
A_\mu^J L^{K}_{ij}f_{JK}^I L_{I}^a.\label{compco3}
\eea 
These new objects appear in the derivatives of the fermions, which
means that
\begin{equation}
\mathcal{D}_{\mu}\lambda^{a}_{i}=\nabla_{\mu}\lambda^{a}_{i}
+\mathcal{Q}_{\mu
i}^{\,\,\,\,\,\,j}\lambda_{j}^{a}+\mathcal{Q}_{\mu}^{ab}
\lambda^{b}_{i}
\end{equation}
should now be understood as containing also the $\gs$ dependent terms
inside the composite connections (\ref{compco1})-(\ref{compco2}).

After these modifications in the supersymmetry transformations
rules
(\ref{trafo2}) and the Lagrangian (\ref{Lagrange2}), the latter
 is supersymmetric up to a small number of
$\gs$ dependent terms. These uncanceled terms take the form
\bea
e^{-1}\,\delta(\cL) & = & \frac12 \gs
\cF^J_{\rho\sigma}f_{JK}^I L_{Iik}L^{Kkj}
\bar\psi_\mu^i\,\Gamma^{\mu\rho\sigma}\varepsilon_j
+\frac{i}{2}\gs\cF^J_{\rho\sigma} f_{JK}^I
L^K_{ij}L_{I}^a\,\bar\lambda^{ia}\Gamma^{\rho\sigma}\varepsilon^j\nnu\\
&& + \gs f_{JI}^K L^{Iij} L_{K}^{a}\cP^{\mu a}_{ij}(\delta A_{\mu}^{J}).
\eea  

\noindent
\textbf{Step 2:}\\
Just as in the Abelian case, the remaining terms can be compensated by
adding fermionic
mass terms as well as potential terms to the Lagrangian, 
and by suitable modifications of the fermionic transformation rules.

The additional mass and potential terms needed are
\bea
\cL_{\textrm{mass}} &=& \frac{3i}{2} \gs
S_{ij}\,\bar\psi_\mu^i\Gamma^{\mu\nu}\psi_\nu^j+i\gs
I_{ijab}\,\bar\lambda^{ia}\lambda^{jb}
+\frac{i}{2}\gs S_{ij}\,\bar\chi^i\chi^j+\gs
T_{ij}^a\,\bar\psi_\mu^i\Gamma^\mu
\lambda^{ja}\nnu\\
&&{}+\sqrt3 \gs S_{ij}\,\bar\psi_\mu^i\Gamma^\mu\chi^j
-\frac{2i}{\sqrt3} \gs T^a_{ij}\,\bar\chi^i\lambda^{ja} -\gs\ga
\cV^{\scriptscriptstyle (AS)} -\gs^2
\cV^\uS,
\eea 
whereas the additional pieces in the transformation rules of the fermions take
the form
\bea
\delta_{\textrm{new}} \psi_{\mu i} &=& i\gs
S_{ij}\Gamma_\mu\varepsilon^j,\\
\delta_{\textrm{new}} \lambda^a_i &=& \gs T_{ij}^a\varepsilon^j,\\
\delta_{\textrm{new}} \chi_i &=& \gs \sqrt3\,S_{ij} \varepsilon^j.
\eea 

The scalar field dependent functions in the above expressions are given by
\bea
S_{ij} &=& -\frac29 \Sigma^{-1} L^{J}_{(i|k|} f_{JI}^K L_{K}^{kl}
L^{I}_{|l|j)},\\
T_{ij}^a &=& -\Sigma^{-1} L^{Ja}L^K_{(i}{}^{k}f_{JK}^I L_{I|k|j)},\\
I_{ijab}&=& -\frac32
S_{ij}\delta_{ab}-\Sigma^{-1}L^{Ja}L^K_{ij}f_{JK}^IL_{I}^b,
\eea
whereas the potential terms read
\begin{equation}
\cV^{\scriptscriptstyle (AS)} \ = \ -18\, U_{ij}\,S^{ij}~,\quad \cV^\uS \ = \
-\frac92\,S_{ij}S^{ij}+\frac12\,T_{ij}^a\,T^{aij}.
\end{equation}

In addition, supersymmetry 
implies a series of derivative relations on the scalar quantities
introduced above, which will be very useful
for the study of the vacua of the theory: 
\bea
\cD_\mu U_{ij} &=&
\frac{2}{\sqrt3}\,\prt_\mu\sigma\,U_{ij}-\frac{2}{3}\,V^a_{(i|k|}\,
\cP_\mu^{ak}{}_{j)},\label{du}\\
\cD_\mu V^a_{ij} &=&  
 2\,N_{[i}^{\ \
  k}{}_{ab}\,\cP_\mu^b{}_{|k|j]}-3 U_{[i|k|}\,\cP_\mu^{ak}{}_{j]}
+\frac{2}{\sqrt3}\,V_{ij}^a\,\prt_\mu\sigma,\label{dv}\\
\cD_\mu S_{ij} &=&
\frac23\,T_{(i|k|}^a\cP_\mu^{ak}{}_{j)}-
\frac{1}{\sqrt3}\,S_{ij}\prt_\mu\sigma,\label{ds}\\
\cD_\mu T_{ij}^a &=& -\frac{1}{\sqrt3}\,\prt_\mu\sigma\,T_{ij}^a+3\,
S_{(i|k|}\cP_\mu^{ak}{}_{j)} -2\,
I_{(i|k|ab}\cP_\mu^{bk}{}_{j)},\label{dt}
\eea
where the derivatives should be understood as being fully $SO(n)$ and
$USp(4)$ covariant derivatives based on the new composite connections and
vielbeins (\ref{compco1})-(\ref{compco3}). Using the numerous constraints
satisfied by the coset representatives $L_\tI^A$, one can show that eqs.\
(\ref{du})-(\ref{dt}) follow automatically from the definitions of the
fermionic shifts.

Just as for the gauging of the Abelian factor, we point out that the shifts in
the supersymmetry laws of the fermionic fields are given by the boosted
structure constants of the gauged semi-simple group.

%%%%%%%%%%%%%%%%%%%%%%%%%%%%%%%%%%%%%%%%%%%%%%%%%%%%%%%%%
\section{The scalar potential $\cV$}
\label{sec_pot}
%%%%%%%%%%%%%%%%%%%%%%%%%%%%%%%%%%%%%%%%%%%%%%%%%%%%%%%%%

The full scalar potential obtained from the $(K_S\times K_A)$ gauging
is
\begin{eqnarray}
\cV \ &=& \ga^2 \cV^\uA + \ga\gs \cV^{\scriptscriptstyle (AS)} 
+ \gs^2 \cV^\uS\nonumber\\
&=& \frac12\,\Big[\; \ga^2\, V_{ij}^aV^{aij} -36\, \ga\gs\, U_{ij}S^{ij}
+\gs^2\,\Big(
 T_{ij}^aT^{aij}-9\, S_{ij}S^{ij}\Big)\,\Big].
\label{KAKS_pot}
\end{eqnarray}

This potential could have been derived directly from the expression of 
the shifts in the supersymmetry laws of the fermionic fields and the 
expression of their kinetic terms.
Indeed, as proved for the first time in four dimensions 
\cite{Cecotti:1986wn}, in all supersymmetric theories the potential 
is given by squaring the shifts of the fermions using the metric 
defined by the kinetic terms:
\begin{equation}
-\frac{1}{2}\, 
\delta_{\varepsilon_1}\bar\psi_\mu^i\,\Gamma^{\mu\nu\rho}\,
\delta_{\varepsilon_2}\psi_{\rho
  i}-\frac{1}{2}\,\delta_{\varepsilon_1}\bar\chi^i\,
\Gamma^\nu\delta_{\varepsilon_2}\chi_i-\frac{1}{2} \delta_{\varepsilon_1}
\bar\la^{ia}\Gamma^\nu\delta_{\varepsilon_2}\la_i^a = \frac{1}{4} 
\bar\varepsilon^i_1 \Gamma^\nu \varepsilon_{2i} \cV,
\end{equation}
provided one satisfies the generalized Ward-identity:
\bea
\frac{1}{4} \Omega_{ij}\,\cV & = &     \frac12 \ga^2 V^a_{i}{}^k V^a_{kj}
 + \ga\gs\Big[9(S_i{}^k U_{kj} +  U_i{}^k
    S_{kj})
+ \frac12 (V^a_{i}{}^k T^a_{kj}-T^a_i{}^k V^a_{kj})\Big]\nnu\\
&&    -\frac12 \gs^2\,\Big[ T^a_i{}^k T^a_{kj} - 9 S_i{}^k S_{kj}\Big].
\eea
This can indeed be verified by using the expression of these objects 
in terms of the coset representatives and then using the properties of 
the latter.

It is instructive to rewrite the scalar potential, as well as the other scalar
field dependent quantities 
in terms of the Killing vectors characterizing the gauged isometries
of the scalar manifold $\cM$. To this end, we switch back to the
parameterization in terms of the $\phi^x$ (cf.\ Sect.\ \ref{gen_set}). 
On the $\phi^x$, $K_S\times K_A$ transformations act as isometries:
\begin{equation}
\delta \phi^x \ = \ \alpha K^x +\alpha^I K_I^x,
\end{equation}
where $\alpha$ and $\alpha^I$ are, respectively, the infinitesimal 
$K_A$ and $K_S$ transformation parameters, whereas $K^x$ and $K^x_I$ denote
the corresponding Killing vectors, which can be expressed as:
\begin{equation}
K^x\ = \ \frac12\,f^{xija}\,\Lambda^M_N\,L_{Mij}L^{Na}~,\quad
K^x_I\ = \ \frac12\,f^{xija}\,L^J_{ij}f^K_{IJ}L_K^a.
\label{kil}
\end{equation}
As in the $\cN=2$ case \cite{CD00,200}, 
one has Killing prepotentials defined as:
\begin{equation}
D_x\cP_{ij}^\uA \ = \ K^y\,\cR_{xyij}~,\quad 
D_x\cP_{Iij}^\uS\ = \ K^y_I\,\cR_{xyij},
\label{prepot}
\end{equation}
where $\cR_{xyij}$ is the $USp(4)$ curvature and the derivatives $D_x$ contain
the original composite connections of the {\em ungauged} theory (see eq.\
(\ref{Q})):
\begin{equation}
Q_x^{ab} \ = \ L^{\tI a}\prt_x
L_\tI{}^b\qquad\mathrm{and}\qquad Q_x^{ij} \ = \
L^{\tI ik} 
\prt_x L_{\tI k}{}^j.
\end{equation}

These prepotentials are found to be:
\begin{equation}
\cP_{ij}^\uA \ = \ -\frac12\,\Lambda^M_N\,L^N_{ik}L_M{}^k_{j},\qquad
\cP_{Iij}^\uS\ = \ \frac12\,f_{IJ}^K L^J_i{}^{k}\,L_{Kkj}~
\label{ppot}
\end{equation}
and are exactly the objects that appear in the shift of the 
composite connection:
\begin{equation}
    \cQ_{\mu i}{}^j = Q_{\mu i}{}^j - 2 \ga a_{\mu} \cP^\uA_{\ \ i}{}^{j} - 2 
    \gs A_\mu^I \cP_{I\ i}^\uS{}^{j}.
\end{equation}
Moreover, the prepotentials have to satisfy the algebra of the $K_S$
isometries \cite{200} 
\begin{equation}
K^x_IK^y_J\,\cR_{xyij}+4 \cP^\uS_{[I\ i}{}^k\cP^\uS_{J]kj}
+f^K_{IJ}\cP^\uS_{Kij} \ = \ 0.
\end{equation}
However, unlike in the $\cN=2$ case \cite{200,CD00}, this does not give any 
additional constraint on the prepotentials. Indeed, 
using eqs.\ (\ref{kil}), (\ref{ppot})
and (\ref{Usp4curv}), one can show that the relation is identically satisfied.

Now, using the above relations, the shifts in the gaugini transformation
rules can be expressed as:
\begin{equation}
V_{ij}^a \ = \ \frac{1}{2\sqrt2}\,\Sigma^2\,f_{xij}^a\,K^x~,\qquad T^a_{ij} \ =
\ \frac12\,\Sigma^{-1}\,L^I_i{}^{k}\,f_{xkj}^a\,K^x_I,
\label{tava_kil}
\end{equation}
whereas 
the shifts in the gravitini transformation rules are given in terms of the
prepotentials:
\begin{equation}
U_{ij} \ = \ \frac{\sqrt2}{3}\,\Sigma^2\,\cP_{ij}^\uA~,\qquad
S_{ij} \ = \ -\frac49\,\Sigma^{-1}\,L^I_i{}^{k}\,\cP_{Ikj}^\uS.
\label{ru_kil}
\end{equation}

A general study of the vacua and domain-wall solutions of this theory is
beyond the scope of this paper, and will be left for a future publication
\cite{DHZ2}.  Instead, we conclude with a list of special
cases, some of which were already studied in the literature.

\begin{itemize}
\item $\gs=0$:

Turning off the $K_S$ gauging leads us back to the Abelian theories studied in
Sect.\ \ref{KA}.
In this case, the potential reduces to\footnote{Note the similarity with the
  corresponding $\cN=2$ theories
\cite{GZ99}, where the part of the scalar potential that is  
due to the presence of tensor multiplets is also non-negative.}    
\begin{equation}
\cV \ = \ \ga^2\,\cV^\uA \ = \ \frac12\,\ga^2\,V^a_{ij} V^{aij} \ = \
\frac12\,\ga^2\,\Sigma^4\,\wh V^a_{ij} \wh V^{aij},
\end{equation}
where, in the following, a hat always denotes the $\sigma$ independent 
part of the scalar field dependent quantities. It
is then easy to see that minimizing the potential with respect to
$\sigma$ requires the potential to vanish at the extrema:
\begin{equation}
\prt_\sigma \cV|_{\phi_c} \ = \ 0 \qquad \Leftrightarrow \qquad \cV|_{\phi_c}
 \ = \ 0.
\end{equation}
Hence, if this potential admits critical points, they have to correspond to
Minkowski ground states of the theory. On the other hand, such critical points
do not necessarily have to exist.
In the case when $K_A$ is {\em compact} (i.e.,
$K_A=U(1)$), however, one can prove that the theory indeed has at least
one Minkowski ground state: expressing the potential in terms of the Killing
vectors, we obtain, using (\ref{tava_kil}):
\begin{equation}
\cV \ = \ \frac14\,\ga^2\,\Sigma^4\,g_{xy}\,K^x K^y.
\label{pa}
\end{equation}
Being compact, the $U(1)$ gauge symmetry 
is a subgroup of the maximal compact subgroup
$\cH=SO(5)\times SO(n)\,$ of $\,\cG=SO(5,n)$, i.e., it is
 a subgroup of the isotropy group of the
scalar manifold. This ensures that there exists at least one
 point $\phi_o\in\cM$
that is invariant under the action of $K_A$, i.e., at this point, 
the $U(1)$ Killing vector and, consequently,
the potential vanish:
\begin{equation}
K^x|_{\phi_o} =0 \quad \Lra \quad \cV|_{\phi_o} =0.
\end{equation}
Hence, there is a least one Minkowski ground state for the $U(1)$ gauged
theory.

\item $\ga=0$:

At a first look, a naive limit $\ga\to 0$ seems to be ill-defined 
due to the presence of the $\frac{1}{\ga}$ term in the Lagrangian
(\ref{Lagrange2}). As was shown in \cite{COW99}, however, there is a
perfectly well-defined procedure to take this limit, based on a field
redefinition of the form
\begin{equation}
B^M_{\mu\nu} \ \lra \ \ga\,C^M_{\mu\nu} + F^M_{\mu\nu},
\end{equation}
where $F^M_{\mu\nu}$ is the curl of some vector field $A_\mu^M$.
After such a limit has been taken, the $K_S\times K_A$ theory 
reduces to a theory of the type discussed in
Sect.\ \ref{sec_KS},
in which only a semi-simple group $K_S$ is gauged.

The scalar potential of such a theory is then given by:
\begin{equation}
\cV \ = \ \gs^2\,\cV^\uS \ = \
\frac12\,\gs^2\,\Big(T^a_{ij}T^{aij}-9\,S_{ij}S^{ij}\Big).
\label{V_S}
\end{equation}
Again, one can factor out the $\sigma$-dependent part, and write the potential
as
\begin{equation}
\cV \ = \ \frac12\,\gs^2\,\Sigma^{-2}\,
\Big(\wh T^a_{ij}\wh T^{aij}-9\,\wh S_{ij}\wh S^{ij}\Big).
\end{equation}
The form of the
$\sigma$-dependence shows that, as in the purely Abelian 
case, at most Minkowski
ground states can exist:
\begin{equation}
\prt_\sigma\cV|_{\phi_c} \ = \ 0 \qquad \Leftrightarrow \qquad \cV|_{\phi_c} 
\ = \ 0.
\end{equation}

However, in contrast to the Abelian case, we cannot prove the existence of
such a vacuum state, even if we assume $K_S$ to be compact.  Indeed, whereas
the $T^a_{ij}$ are expressed in terms of the Killing vectors as in eq.\ 
(\ref{tava_kil}), the $S_{ij}$ are proportional to the Killing prepotentials,
which need not vanish at the invariant point $\phi_o\in\cM$ where the Killing
vectors are zero for a compact $K_S$.  Eq.\ (\ref{prepot}) merely implies that
the prepotentials should be covariantly constant at that point.

The particular choice $K_S=SU(2)\subset SO(5)$ corresponds to the case studied
in \cite{AT85}. Indeed, our potential (\ref{V_S}) has precisely the same form
as the potential of ref.\ \cite{AT85}. However, in our case, we have
\begin{equation}
\prt_x\cV \ \neq \ 0,
\end{equation}
due to the fact that we allowed for a general mass term $I_{ijab}$ 
for the gaugini, that also contains a part antisymmetric in $a,b$, whereas in
\cite{AT85}, one has $\;I_{ijab}=-\smf32 S_{ij}\delta_{ab}$. Thus, it seems 
that the $SU(2)$ gauging considered in \cite{AT85} is not the most general one.

\item $n=0$:
  
  In this case, only the $\cN=4$ supergravity multiplet is present.
  This means that the global symmetry group $SO(1,1)\times SO(5,n)$ of the
  ungauged theory (\ref{Lagrange1}) reduces to $SO(1,1)\times SO(5)$, where
  $SO(5)$ is just the $R$-symmetry group of the $\cN=4$ Poincar\'e
  superalgebra. As discussed in Section \ref{sec_KS}, $K_S$ can then only be
  the standard $SO(3)\cong SU(2)$ subgroup of $SO(5)$, and we therefore expect
  to recover the $SU(2)\times U(1)$ gauged theory of \cite{RO86} when this
  $SU(2)$ is gauged together with the obvious additional $SO(2)\cong U(1)$
  subgroup of $SO(5)$.
  
  Indeed, in the absence of exterior vector multiplets, we have
  $\;V^a_{ij}=T^a_{ij}=0$, whereas the $\sigma$ independent quantities $\;\wh
  U_{ij}$ and $\wh S_{ij}$ become constant matrices ($\wh U_{ij} = \frac{1}{6}
  (\Gamma_{45})_{ij}$ and $\wh S_{ij} = -\frac{2}{9} (\Gamma_{123})_{ij} = -
  \frac{2}{9} (\Gamma_{45})_{ij}$)\footnote{The gamma matrices introduced here
    refer to Euclidean gamma matrices of $SO(5)$.}  such that the potential
  (\ref{KAKS_pot}) reduces to
  \begin{equation} 
      \cV \ = \ -\gs\,\Big(\ga\,\Sigma+\gs\,\Sigma^{-2}\Big), 
  \end{equation}  
  where we have discarded some numerical factors.  This potential indeed
  coincides with the potential of \cite{RO86}.
  
\item $n=2, K=SU(2)\times U(1)$:

In the ungauged theory with scalar manifold
\begin{equation}
\cM \ = \ \frac{SO(5,2)}{SO(5)\times SO(2)}\times SO(1,1),
\end{equation}
one can gauge an $SU(2)\times U(1)$ subgroup such that four tensor fields
appear, two in the supergravity multiplet and two coming from the former
vector multiplets. To this end, the $SU(2)$ factor has to be the standard
$SO(3)\cong SU(2)\subset SO(5)$, whereas the $U(1)$ factor has to be a
diagonal subgroup of the two $SO(2)$'s in the obvious subgroup $SO(2)\times
SO(3)\times SO(2)\subset SO(5,2)$.  This corresponds to the $\cN=4$ theory
obtained in \cite{FGPW99} 
by truncating the $\cN=8$ theory to an $SU(2)_I\subset SU(4)$
invariant subsector. The relevance of this
sector for the study of certain RG-flows in the AdS/CFT correspondence was
stressed in the Introduction.
\end{itemize}

\section{Conclusions}

In this paper, we have studied the possible gaugings of matter coupled $\cN=4$
supergravity in five dimensions. All these theories can be obtained from the
ungauged MESGTs of ref.\ \cite{AT85} by gauging appropriate subgroups of the
global symmetry group $SO(5,n)\times SO(1,1)$, where $n$ is the number of
vector plus tensor multiplets. A more careful analysis of the possible gauge
groups revealed that the $SO(1,1)$ factor cannot be gauged, whereas the
possible gaugeable subgroups $K$ of $SO(5,n)$ naturally fall into three
different categories:

If $K$ is Abelian, its gauging requires the dualization of the $K$ charged
vector fields into self-dual antisymmetric tensor fields.  For consistency
with the structure of the ungauged theory, such an Abelian gauge group $K$ has
to be one-dimensional, i.e., it can be either $U(1)$ or $SO(1,1)$. In the case
$K=U(1)$, we could prove the existence of at least one Minkowski ground state.
Such an existence proof could not be given for the case $K=SO(1,1)$
\footnote{In fact, in \cite{GZ00} a counter-example was found for a particular
  non-compact gauging of the $\cN=2$ analog.}. In any case, however, no AdS
ground states are possible if $K$ is Abelian.

The gauging of a semi-simple group $K\subset SO(5,n)$, by contrast, does {\em
  not} introduce any tensor fields. Conversely, self-dual tensor fields can
only be charged with respect to a one-dimensional Abelian group. This is an
interesting difference to the $\cN=2$ and $\cN=8$ cases
\cite{GRW86,PPN85,GZ99}, where the tensor fields could also be charged under a
non-Abelian group. As for the critical points of the scalar potential, we
found that they can at most correspond to Minkowski space-times. However, we
were not able to give an existence proof similar to the $U(1)$ case. To sum
up, neither the pure Abelian gauging nor the gauging of a semi-simple group
alone allow for the existence of AdS ground states. As for $\cN=4$
supersymmetric AdS ground states, this is to be expected from the
corresponding AdS superalgebra $SU(2,2|2)$, which has the $R$-symmetry group
$SU(2)\times U(1)$.

This situation changes when $K=K_S\times K_A$ is the direct product of a
semi-simple and an Abelian group. In this case, the Abelian group $K_A$ again
has to be one-dimensional, and tensor fields are required for its gauging. The
resulting scalar potential for the simultaneous gauging of $K_S$ and $K_A$
consists of the sum of the potentials that already arise in the separate
gaugings of $K_S$ and $K_A$, but also contains an interference term that
depends on both coupling constants. Again, this is an interesting difference
to the analogous $\cN=2$ gaugings \cite{GZ99}, where no such interference was
found. Due to all these different contributions to the scalar potential, it is
now no longer excluded that AdS ground states exist. In fact, this is to be
expected from various special cases that have appeared in the literature
\cite{RO86,FGPW99}. On the other hand, the more complicated form of this
potential also makes a thorough analysis of its critical points along the
lines of, e.g., \cite{041} more difficult.

At this point, we should mention that the conclusions of our analysis on the
possible gaugings might be modified if one were to consider even more general
gaugings based on non semi-simple algebras as was done, e.g., in the $\cN=8$
case in \cite{ACFG}.

Having obtained the general form of matter coupled $\cN=4$ gauged
supergravity, one might now try to clarify several open problems.  First, it
would be interesting to see whether some of these theories can indeed be used
to extract information on the $\cN=2$ quiver gauge theories discussed in
\cite{DM96,LNV98,403,G98,FGPW99,411}, and to recover the gravity duals of
various RG-flows that have been constructed \cite{FGPW99} or were conjectured
to exist \cite{KW98}.  Furthermore, one now has the tools to elucidate some of
the questions raised in \cite{BHLT00} concerning possible alternative gaugings
of certain $\cN=2$ matter coupled theories.

Another interesting issue would be to clarify whether the gaugings presented
here can be reproduced by dimensional reduction of the heterotic string on a
torus with internal fluxes, as was done in \cite{KM99}. 
In fact, the dimensional reduction does not introduce any additional 
tensor fields (beside the NS-NS two-form). As these were shown to play a
essential r{\^o}le for the Abelian gauging, one might expect some
differences concerning this part of the gauging between the two approaches.

Finally, it should be extremely interesting to use this new theory to obtain
new insight on the existence of smooth supersymmetric brane-world solutions
\`a la Randall-Sundrum.

Even if, so far, the study of the $\cN = 2$ theory has not yet provided a
final answer \cite{proc}, it has anyhow restricted the analysis to models
involving vector- and hypermultiplets.  Unfortunately, there are many possible
scalar manifolds and gaugings that one can build for such models.  In any
case, the models that can be obtained by reduction of some $\cN = 4$
realization must be contained in the general setup presented here.  Moreover,
since the $\cN = 4$ scalar manifold has the unique form presented in \eqn{cM},
the analysis of the possible gaugings and flows, should be simpler and we hope
that it should lead to more stringent results.

\bigskip

\textbf{Acknowledgments:} The work of G.D.\ is supported by the European
Commission under TMR project HPRN-CT-2000-00131.  The work of C.H.\ is
supported by the German-Israeli Foundation for Scientific Research (GIF).
The work of M.Z.\ is supported by the German Science Foundation (DFG) within
the Schwerpunktprogramm 1096 ``Stringtheorie''.
\vspace{0.5cm}

\noindent
{\bf\Large Appendix}

\begin{appendix}
\section{Spacetime and gamma matrix conventions}

This appendix summarizes our spacetime and 
gamma matrix conventions. 
The spacetime metric $g_{\mu\nu}$ and the f\"{u}nfbein $e_{\mu}{}^{m}$
are related by
\begin{displaymath}
g_{\mu\nu}=e_{\mu}{}^{m}e_{\nu}{}^{n}\eta_{mn}
\end{displaymath}
with $\eta_{mn}=\textrm{diag}(-++++)$. The indices $\mu,\nu\ldots$
and $m,n,\ldots$ are `curved' and `flat' indices, respectively,
and run from $0$ to $4$. Our conventions regarding the Riemann tensor
and its contractions are
\bea
R_{\mu\nu mn}(\omega)&=&[(\prt_{\mu}\omega_{\nu mn} + \omega_{\mu mp}\,
\omega_{\nu}{}^{p}{}_{n})-(\mu\leftrightarrow \nu)],\\
R(\omega,e)&=&e^{\nu m}e^{\mu n}R_{\mu\nu mn},
\eea
where $\omega_{\mu mn}(e)$ is  the  spin connection defined via the
usual constraint
$$\prt_{[\mu}e_{\nu]m}+\omega_{[\mu m}{}^{n}e_{\nu ]n}\ = \ 0.$$
The gamma matrices $\Gamma_{m}$ in five dimensions
 are constant, complex-valued 
$(4\times 4)$-matrices satisfying
\begin{displaymath}
\{\Gamma_{m},\Gamma_{n}\}=2\eta_{mn}.
\end{displaymath}
Given a set $\{\Gamma_{0},\Gamma_{1},\Gamma_{2},\Gamma_{3}\}$ of gamma 
matrices in \emph{four} dimensions, the definition 
\begin{equation}
\Gamma_{4}:=\pm i\Gamma_{0}\Gamma_{1}\Gamma_{2}\Gamma_{3} 
\end{equation}
yields a representation of the corresponding Clifford algebra in five 
dimensions for either choice of the sign. We will always choose   this 
sign such that eq.\ (\ref{Gamma=iepsilon}) below holds.
The gamma matrices $\Gamma_{\mu}$ with a `curved' index are simply
defined by
$\Gamma_{\mu}:=e_{\mu}{}^{m}\Gamma_{m}$. 

Antisymmetrized products of gamma matrices are denoted by
\begin{displaymath}
\Gamma_{\mu_{1}\mu_{2}\ldots\mu_{p}}:=\Gamma_{[\mu_{1}}\Gamma_{\mu_{2}}\ldots
\Gamma_{\mu_{p}]},
\end{displaymath}
where the square brackets again denote antisymmetrization with 
``strength one''. Using this definition, we have
\begin{equation}
\Gamma^{\mu\nu\rho\sigma\lambda} \ = \
\frac{i}{e}\,\epsilon^{\mu\nu\rho\sigma\lambda}.
\label{Gamma=iepsilon}
\end{equation}
The charge conjugation matrix $C$ in five dimensions satisfies
\begin{equation}\label{CGamma}
C\Gamma^{\mu} C^{-1}=(\Gamma^{\mu})^{T}.
\end{equation}
It can be chosen such that
\begin{displaymath}
C^{T}=-C=C^{-1}.
\end{displaymath}
The antisymmetry of $C$ and the defining property (\ref{CGamma}) imply that
the matrix $(C\Gamma_{\mu_{1}\ldots\mu_{p}})$ is symmetric for $p=2,3$ and
antisymmetric for $p=0,1,4,5$. 

All spinors we consider are anticommuting 
symplectic Majorana spinors
\begin{displaymath}
\bar{\chi}^{i}:=(\chi_{i})^{\dagger}\Gamma_{0}=\Omega^{ij}\chi^{T}_{j}C.
\end{displaymath}
As a consequence of the symmetry properties of 
$(C\Gamma_{\mu_{1}\ldots\mu_{p}})$, these (anticommuting) spinors satisfy 
the following useful identities:
\begin{displaymath}
{\bar{\Psi}}^{i}\Gamma_{\mu_{1}\ldots\mu_{p}}\chi^{j}=\left\{
\begin{array}{ll}
+{\bar{\chi}}^{j}\Gamma_{\mu_{1}\ldots\mu_{p}}\Psi^{i}& (p=0,1,4,5),\\
-{\bar{\chi}}^{j}\Gamma_{\mu_{1}\ldots\mu_{p}}\Psi^{i}& (p=2,3).
\end{array}
\right.
\end{displaymath}
which implies
\begin{displaymath}
{\bar{\Psi}}^{i}\Gamma_{\mu_{1}\ldots\mu_{p}}\chi_{i}=\left\{
\begin{array}{ll}
-{\bar{\chi}}^{i}\Gamma_{\mu_{1}\ldots\mu_{p}}\Psi_{i}& (p=0,1,4,5),\\
+{\bar{\chi}}^{i}\Gamma_{\mu_{1}\ldots\mu_{p}}\Psi_{i}& (p=2,3).
\end{array}
\right.
\end{displaymath}

\section{The geometry of the scalar manifold}

The scalars of the theories studied in this paper 
parameterize the manifold:
\begin{equation}
\label{coset}
\cM \ = \ \frac{SO(5,n)}{SO(5)\times SO(n)}\times SO(1,1)
\end{equation}
where the $SO(1,1)$ factor is spanned by the scalar $\sigma$ 
of the supergravity 
multiplet  and the 
rest by the scalars $\phi^x$ of the matter multiplets.

Having introduced the coset representatives $L_{\tilde I}{}^A$, one can
determine the metric and curvatures of the manifold \eqn{coset}. Indeed the
vielbeins are obtained from
\begin{equation}
\label{vielbein}
f_{x}^{aij} = - 2 L^{\tilde{I} a} \partial_{x} L_{\tilde I}^{ij},
\end{equation}
whereas the metric is given by
\begin{equation}
f_{x}^{ija} \, f_{y\,ij}^a = 4 g_{xy}.     
\label{metric}
\end{equation}
The inverse of the vielbeins are defined via
\begin{equation}
f_{x}^{ija} f^{x\,b}_{kl} = 4 
\Big( \delta^{[i}_{k}\delta^{j]}_{l} - \frac{1}{4} 
\Omega^{ij} \Omega_{kl}\Big) \delta^{ab}.
\label{idf}
\end{equation}
As usual, the vielbeins are covariantly constant with respect to the full
covariant derivative:
\begin{equation}
    \partial_{x} f_{y}^{ij a} - {\Gamma_{xy}}^z f_{z}^{ij a} -
    {Q_{x\,b}}^a f_{y}^{ij b}  - {Q_{x\,k}}^i f_{y}^{kj a} - {Q_{x\,k}}^j
    f_{y}^{ik a} = 
    0,
    \label{covcost}
\end{equation}
where $\Gamma_{xy}^z$ denotes the Christoffel symbols on $\cM$.

The $USp(4)$ and $SO(n)$ curvatures fulfill the following identities:
\begin{eqnarray}
{R_{xy\,i}}^j & = & -\frac{1}{4} f_{[x\,ik}^a \, f_{y]}^{a\,kj},
\label{Usp4curv}  \\
{R_{xy\,a}}^b & = & \frac{1}{4} f_{[x\,a}^{ij} \, f_{y]\,ij}{}^b,
\label{SOncurv}
\end{eqnarray}
which can also be found as the solutions to the 
 integrability conditions coming 
from the differential equations satisfied by the coset representatives
\begin{eqnarray}
    D_x L_{\tI ij} & = & -\frac{1}{2} L_{\tI}^a f_{xij}^a,\label{diff1}  \\
    D_x L^{\tI}_{ij} & = & \frac{1}{2} L^{\tI\,a} f_{xij}^a,\label{diff2}  \\
    D_x L^a_{\tI} & = & -\frac{1}{2} f_{xij}^a L_{\tI}^{ij}, \label{diff3} \\
    D_x L^{\tI a} & = & \frac{1}{2} f_{x}^{ij a} L^{\tI}_{ij},\label{diff4}
\end{eqnarray}
where $D_x$ denotes the $USp(4)$ and $SO(n)$ covariant derivative.
Let us finally display the useful identity:
\begin{equation}
L_{(\tI ik}L_{\tJ)}{}^{jk} \ = \ \frac14\,\delta_i^j\,L_\tI{}^{kl}L_{\tJ kl},
\end{equation}
which is nothing but the defining relation of the $SO(5)$
Clifford algebra used to convert 
the $SO(5)$ index of the $L_\tI{}^A$ into
the corresponding composite $USp(4)$ index with the property that
\begin{equation}
L_\tI{}^{ij} \ = \ -L_\tI{}^{ji}~,\qquad  L_\tI{}^{ij}\,\Omega_{ij} \ = \ 0.
\end{equation}

\section{Superspace analysis}

In this appendix we want to propose a brief analysis of the  
constraints needed to reproduce the theory built in the main text in 
superspace.
This will help us understand to which extent the couplings proposed 
here (besides non-minimal ones) are general.

In the superspace formalism, the dynamics of the fields is determined 
by the constraints one imposes on the supercurvatures.
To classify the possible consistent sets of constraints, one adopts a 
general strategy which is based on group-theoretical analysis \cite{CL}.
The superspace formalism gives quite stringent restrictions 
on the possible couplings of the various multiplets of the theory and
the analysis of the lowest-dimensional components of the various
superfields (supertorsion and supercurvatures) reveals what freedom 
one has in coupling for instance the gravity multiplet to the matter ones.

Moreover, we report some additional equalities, derived by the solution of the 
superspace Bianchi identities for the various supercurvatures, which 
are useful to derive some of the properties of the shifts in the 
transformation laws of the fermionic fields.
The same relations could be derived at the component level by closing 
the supersymmetry algebra on the various fields.

\bigskip

We denote a  generic superform as
\begin{equation}
    \Phi = \frac{1}{n!} e^{A_1} \ldots e^{A_n} \Phi_{A_n \ldots A_1},
\end{equation}
where $A=(\mu,\alpha i)$ collectively denote vector ($\mu$) and spinor
 ($\alpha i$) indices, 
and $e^A$ are the supervielbeins
(which means that the projection of $e^{\a i}$ on ordinary space-time 
$dx^\mu \psi_\mu^{\a i}$ contains the gravitino field). 

The (super-)torsion and (super-)curvatures are defined as 
\begin{eqnarray}
D e^A = T^A, && D T^A = e^B {R_B}^A, \\
F^I= dA^I +\frac12\,\gs f^I_{JK} A^J A^K, && G = d a,\\
H^M = d B^M + \ga \Lambda^M_N  \, a B^N,
\end{eqnarray}
and they satisfy the Bianchi identities:
\begin{eqnarray}
D T^A = e^B {R_B}^A, && DR_{B}{}^A = 0,\\
DF^I= 0, && d G = 0,\\
D H^M = \ga \Lambda^M_N  \, G B^N.
\end{eqnarray}
At this level, one tries to impose constraints on such superfields which are compatible
with their Bianchi Identities (BI) such as to remove all the auxiliary fields
that appear in their component expansion.

In the language of superspace, the basic constraint one has to impose to
close $\cN = 4$ supersymmetry in a linear way is given by
\begin{equation}
T^m_{\a i \b j} = \frac{1}{2} \Omega_{ij} \Gamma^m_{\a\b}.
\end{equation}

The other basic constraints on the torsion field are given by the definition
of the ${T_{\a i m}}^n$ and ${T_{\a i \b j}}^{\gamma k}$ components.
To analyze the freedom we have, and therefore the possibility of coupling
gravity with matter, we make a group theoretical analysis of these structures.

The  ${T_{\a i \b j}}^{\gamma k}$ tensor contains the following $SO(5)_{USp(4)_R}$
representations
$ 5 \times {\bf 4_4} + 3 \times {\bf 4_{16}} + {\bf 4_{20}} + 3 \times {\bf
16_4} + 2 \times {\bf 16_{16}} + {\bf 16_{20}} + {\bf 20_4} + {\bf 20_{16}}
+ {\bf 20_{20}}$, whereas  ${T_{\a i m}}^n$ contains
$ 2 \times {\bf 4_4} + 2 \times {\bf 16_4} + {\bf 20_4} + {\bf 40_4}$.

Using the connection redefinition $\omega_{\a i\, m}{}^n \to \omega_{\a i\,
  m}{}^n + X_{\a i\, [ m p]} \eta^{pn}$ we are free to reabsorb many of the
components of ${T_{\a i m}}^n$ which then remains with only the ${\bf 4_4} +
{\bf 16_4} + {\bf 40_4}$ representations.  Moreover, using the gravitino
redefinition $\psi^{\a i} \to e^m H_m{}^{\a i}$ we are left with only the
${\bf 40_4}$.  This latter is then set to zero by the lowest dimensional
$T$-BI, which is the equation $R_{(\a i \b j \gamma k)}{}^m = 0$.

At the same time the ${T_{\a i \b j}}^{\gamma k}$ tensor remains with the
irreps $2 \times {\bf 4_4} + {\bf 4_{16}} + {\bf 4_{20}}$ which allows only
for structures of the kind $\chi_{\a i} \delta_\b^\gamma \Omega_{jk}$ and
$Q_{\a i \,(jk)} \delta_\b^\gamma$.  Indeed the first kind of term is needed
to close the BI of the graviphoton, whereas the second is used to couple
matter (it has the same tensor structures as an $USp(4)$ connection, or more
precisely its pull-back on superspace).  This means that, even if one has not
yet defined the scalar manifold, the type of interaction between the gravity
sector and the matter must be mediated by the appearance of some type of
$USp(4)$ connection field.

Once the connection-type term has been reabsorbed in the definition of a new
supercovariant derivative and the $F$-BI (with coupling to matter) have been
solved, one finds that the definition of the torsion constraint which is
compatible with the supersymmetry transformations presented in this paper is
given by
\bea
{T_{\a i \b j}}^{\gamma k} &= & \frac{i}{2\sqrt{3}} \left( \chi_{\a i} 
\delta_{\b j}^{\gamma k} + \chi_{\b j} \delta_{\a i}^{\gamma k} + 
2 \chi^{\gamma k} C_{\a\b} \Omega_{ij} + 4 \chi^\gamma_{[i} \delta_{j]}^k 
C_{\a\b} 
-  \chi_{\a j} \delta_{\b i}^{\gamma k} \right.\nonumber \\
&& \phantom{\frac{i}{2\sqrt{3}} \left(\right.}
 - \left. \chi_{\b i} \delta_{\a j}^{\gamma k} 
+ \chi_\a^k \delta_\b^\gamma \Omega_{ij} - 
\chi_\b^k \delta^\gamma_\a \Omega_{ij}\right),
\eea
which translates in components as the following three-Fermi term in the 
transformation rule
of the gravitino:
\bea
\delta_{\varepsilon} \psi_\mu^k &=& \ldots + \frac{i}{2\sqrt{3}} 
\left( \bar{\psi}_\mu^i 
\chi_i \varepsilon^k - \bar{\varepsilon}^i \chi_i \psi_\mu^k + 2 \bar{\psi}_\mu^i\varepsilon_i \chi^k - 2
    \bar{\psi}_\mu^k \varepsilon^i \chi_i \right. \nonumber\\
& &\phantom{\ldots + \frac{i}{2\sqrt{3}} 
\left(\right.}+ \left. 2\,\bar{\psi}_\mu^i \varepsilon^k \chi_i + \bar{\varepsilon}^k \chi_i
    \psi_\mu^i - \bar{\psi}_\mu^k \chi_i \varepsilon^i - \bar{\varepsilon}_i \chi^k \psi_\mu^i -
    \bar{\psi}_\mu^i \chi^k \varepsilon_i\right),
\eea
where in the dots are the components presented in the main text and 
the three-Fermi terms due to the pull-back of the $USp(4)$ 
connection.

For completeness, we list here also the other fundamental 
constraints:
\bea
D_{\a i} \si &=& -\frac{i}{2} \chi_{\a i}, \\
D_{\a i} \phi^{x} &=& -\frac{i}{2} \lambda_{\a}^{j a} f^{x}_{ij\,a}, \\
G_{\a i\b j} &=& - \frac{i}{2\sqrt{2}} \Sigma^2 \Omega_{ij} C_{\a\b}, \\
F^I_{\a i\b j} &=& i \Sigma^{-1} L^I_{ij} C_{\a\b}, \\
H^M_{\a i\b j\g k} &=& 0.
\eea

As promised, we also show the equalities that one derives between the 
various shifts in the supersymmetry laws of the Fermi fields by 
solving the superspace Bianchi identities (BI).
From the $H$-BI we derive
\begin{equation}
\label{cond1}
(-4 U + 2 S)_{[i}{}^k L_{j]k}^M - V^{a}_{[ij]} L_{a}^M + 2
(S+U)_{[i}{}^k L_{j]k}^M = \Sigma^2 \frac{1}{\sqrt{2}} \Lambda^M{}_N 
L^N_{ij},
\end{equation}
whereas from the $F$-BI we get
\begin{equation}
\label{cond2}
(-4U +2 S)_{[i}{}^k L_{j]k}^M + V^{a}_{[ij]} L_{a}^M + 2
(S+U)_{[i}{}^k L_{j]k}^M = 0.
\end{equation}
Moreover, the closure of the $\phi$-BI implies that 
\begin{eqnarray}
V^{a}_{ij} &=& \frac{1}{\sqrt{2}} \, \Sigma^2 \, \Lambda^N{}_M L_{N
ij}L^{Ma} \\
T^a_{ij}&=&  -\Sigma^{-1} \, f^I_{JK} \, L_{I (i}{}^m L^J_{|m|j)} 
L^{Ka}.
\end{eqnarray}

Using the solution of these quantities in terms of the coset 
representatives, the \eqn{cond2} equation becomes just $C^{IM} = 0$ (for the
Abelian part, an identity for the non-Abelian).

\end{appendix}

\providecommand{\href}[2]{#2}

\end{document}